# Dynamic Estimation of Credit Rating Transition Probabilities

**Arthur M. Berd**
Lehman Brothers Inc.

*We present a continuous-time maximum likelihood estimation methodology for credit rating transition probabilities, taking into account the presence of censored data. We perform rolling estimates of the transition matrices with exponential time weighting with varying horizons and discuss the underlying dynamics of transition generator matrices in the long-term and short-term estimation horizons.*

## INTRODUCTION

Despite the advances of market-implied models of credit risk such as those developed by Moody's KMV (Crosbie and Bohn [2001], Kealhofer [2003]) and RiskMetrics' CreditGrades (Finkelstein, Pan, Lardy, Ta and Thierney [2002]), credit ratings remain a very influential signal in both investment grade and high yield markets. While the rating changes may indeed lag the market developments, the information contained in the act of a downgrade or upgrade as well as the specific language used to characterize its reasons and further credit outlook remains at the forefront of investors' minds and retains some explanatory power even when combined with structural models of credit risk (see Sobehart, Stein, Mikityanskaya and Li [2000] and Sobehart, Stein and Keenan [2000]).

The aggregate market-wide rating transition trends, such as upgrade/downgrade ratios, as well as the realized default rates (we consider default as a particular form of rating transition), are widely followed by portfolio managers as a gauge of the credit cycle. Many long-term credit investors, including those managing bank loan portfolios, insurance assets and CDOs, often pose a question: "what is the current estimate of the N-year rating transition probability?" (where the typical values of N can be 1, 5 or 10 years). Robust estimates, and, if possible, forecasts of rating transition risks are important ingredients of their macro investment strategies. In this article we develop a methodology that addresses such needs.

The paper is organized as follows. First we formally define the continuous-time Markov description of the rating transition process. We give a brief overview of the commonly used cohort estimators that are reported by rating agencies and followed by many practitioners and explain their shortcomings. Next, following the ideas developed by Lando *et al.* (Lando [2000], Lando and Skodeberg [2004], Christensen, Hansen and Lando [2004] and Fidelius, Lando and Nielsen [2004]), we define a maximum likelihood estimator for the generator matrix of transitions. We then extend these academic models to allow for time-weighted estimation procedures and for parametric smoothing of the generator matrix, which provide a simultaneously dynamic and robust practical solution to the question above.

Finally, we present the results of the rolling estimation of the transition probabilities corresponding to different levels of the exponential decay rate, and examine whether such estimates of the "recent" transition matrices can be used as forecasts of the future rating transition risk. In particular, we determine the optimal choice of the half-life (decay) parameter, which maximizes the out-of-sample forecasting power of the time-weighted estimates given the forecast horizon. Not surprisingly, we conclude that there is a significant degree of inertia in rating transitions, and that the shorter forecast horizons require a shorter half-life, whose optimal value is close to the target horizon.

## A CONTINUOUS-TIME MARKOV MODEL OF RATING TRANSITIONS

Let us assume that at any time the credit risk of each issuer is fully determined by its rating, belonging to a finite set with K ratings, for example an eight-state letter-grade rating system [Aaa, Aa, …, Caa, D] or a eighteen-state notched rating system [Aaa, Aa1, Aa2, …, Caa, D]. Note that we explicitly count default as the last $K$-th state.

The period-$(0,t)$ transition matrix is defined as the set of probabilities of finding the final rating of a company at state $q$ (possibly default) at time $t$, under the condition that the initial rating was $r$ at time $t=0$. The transition process is assumed to be Markov, but not necessarily stationary, i.e. the compound transition probability from time $t=0$ to time $t=s$ is obtained by summation over all possible intermediate states at time $t=u$. This can be described as a matrix multiplication of the transition matrix for period $(0,u)$ and for period $(u,s)$, respectively[1].

$$[1] \qquad T_{rq}(0,s) = \sum_p T_{rp}(0,u) \cdot T_{pq}(u,s)$$

This relationship is illustrated in Figure 1, where we depict various paths, which must be summed over to arrive at the transition probability from initial rating to the final one:

**Figure 1.    Markov dynamics of credit ratings**

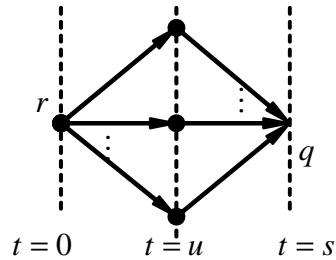

For the probabilistic interpretation of the transition matrix $T_{rq}$ to be possible, all of its elements must be positive, and each row must sum up to 1. One typically assumes that the default is absorbing state, i.e. there are transitions into it but not out of it. In terms of transition matrices, this means that the elements of the last row of the matrix are zero everywhere except for the diagonal, which is equal to one.

For stationary transition probabilities, one can introduce the so-called generator matrix, using the notions of matrix exponential and matrix logarithm:

$$[2] \qquad T(t) = \text{mexp}(t \cdot \Lambda), \text{ and } \quad \Lambda = \frac{1}{t} \cdot \text{mlog}(T(t))$$

Consider a limit of small time intervals:

$$[3] \qquad T_{rq}(\Delta t) = \mathrm{I}_{rq} + \Delta t \cdot \Lambda_{rq}$$

---

[1]    The order of matrix multiplication is related to the definition of the transition matrix. If the transition matrix is defined so that the initial rating corresponds to the rows and final to the columns (therefore the sum of probabilities along each row is 1), then the matrix multiplication should be done left-to-right. In the alternative case when the initial rating is along the columns and final is along the rows, the matrix multiplication should be done right-to-left. We use the former convention is this paper.

Since the transition matrix must have non-negative elements, and the identity matrix I has zeros off-diagonal, it follows that the generator matrix must have non-negative elements off-diagonal. Furthermore, since the rows of the transition matrix add up to 1, and the rows of the identity matrix also add up to 1, it follows that the rows of the generator matrix must add up to 0. In other words, the diagonal element of the generator matrix is equal to the negative of the sum of its off-diagonal elements.

[4] $$\lambda_{rr} = -\sum_{q \neq r} \lambda_{rq}$$

Under the absorbing default state assumption, all elements of the last row of the generator matrix are equal to zero.

Examples of a stationary transition probability matrix and a generator matrix are shown in Figures 2 and 3. Figure 2 shows a smoothed 1-year transition probability matrix based on a long-term (20+ years) historical Moody's estimate, as used in the popular CreditMetrics model (see Gupton, Finger, Bhatia [1997]). The rows denote the initial ratings, the columns refer to the final ratings. Note that the last (default) row contains all zeros off-diagonal.

Figure 3 shows the generator matrix that was derived from the given transition probability matrix, assuming stationarity. We have taken the matrix log of the transition matrix, as in [2], and had to override a few entries (shown in bold) to avoid negative numbers. Note that the diagonal elements are all negative (highlighted in red) and that the last row is zero.

One can also confirm by direct computation that the sum of elements on each row of the transition matrix in Figure 2 is equal to 1, and the sum of elements on each row of the generator matrix in Figure 3 is equal to zero, in accordance with eq. [4].

**Figure 2.    One-year rating transition probability matrix**

| Rating | Aaa | Aa | A | Baa | Ba | B | Caa | D |
|---|---|---|---|---|---|---|---|---|
| Aaa | 0.8812 | 0.1029 | 0.0102 | 0.0050 | 0.0003 | 0.0002 | 0.0001 | 0.0001 |
| Aa | 0.0108 | 0.8870 | 0.0955 | 0.0034 | 0.0015 | 0.0010 | 0.0004 | 0.0003 |
| A | 0.0006 | 0.0288 | 0.9011 | 0.0592 | 0.0074 | 0.0016 | 0.0006 | 0.0008 |
| Baa | 0.0005 | 0.0034 | 0.0707 | 0.8504 | 0.0605 | 0.0101 | 0.0028 | 0.0016 |
| Ba | 0.0003 | 0.0008 | 0.0056 | 0.0568 | 0.7957 | 0.0808 | 0.0454 | 0.0146 |
| B | 0.0001 | 0.0004 | 0.0017 | 0.0065 | 0.0659 | 0.8270 | 0.0276 | 0.0706 |
| Caa | 0.0001 | 0.0002 | 0.0064 | 0.0105 | 0.0305 | 0.0611 | 0.6296 | 0.2616 |
| D | 0 | 0 | 0 | 0 | 0 | 0 | 0 | 1 |

*Source: RiskMetrics Group and Lehman Brothers*

**Figure 3.    Transition generator matrix derived from transition probability matrix**

| Rating | Aaa | Aa | A | Baa | Ba | B | Caa | D |
|---|---|---|---|---|---|---|---|---|
| Aaa | -0.1272 | 0.1164 | 0.0050 | 0.0055 | 0.0001 | 0.0001 | 0.0001 | 0.0001 |
| Aa | 0.0122 | -0.1223 | 0.1070 | 0.0002 | 0.0012 | 0.0010 | 0.0005 | 0.0002 |
| A | 0.0005 | 0.0321 | -0.1086 | 0.0676 | 0.0062 | 0.0011 | 0.0005 | 0.0007 |
| Baa | 0.0006 | 0.0025 | 0.0807 | -0.1674 | 0.0732 | 0.0084 | 0.0013 | 0.0008 |
| Ba | 0.0004 | 0.0007 | 0.0035 | 0.0685 | -0.2363 | 0.0973 | 0.0622 | 0.0037 |
| B | 0.0001 | 0.0004 | 0.0015 | 0.0048 | 0.0807 | -0.1952 | 0.0356 | 0.0721 |
| Caa | 0.0001 | 0.0001 | 0.0078 | 0.0124 | 0.0391 | 0.0825 | -0.4657 | 0.3238 |
| D | 0 | 0 | 0 | 0 | 0 | 0 | 0 | 0 |

*Source: Lehman Brothers*

In a special case when the rating transitions are possible only directly between the initial rating and the default state, and assuming an absorbing default state, we are left with a generator matrix whose only non-zero elements are in its final (default) column and on its diagonal (with zero final row). The matrix exponential [2] in this case acquires a shape with non-zero elements on its diagonal, representing the survival probability for an issuer of rating q until time t, the last column representing the corresponding cumulative default probability, and all other elements being zero. Thus, in this case we recover the usual relationship between the survival and cumulative default probabilities and the forward hazard rate:

$$[5] \quad Q_r(0,t) = T_{rr}(0,t) = \exp\left(-\int_0^t ds \cdot h_r(s)\right)$$

$$[6] \quad D_r(0,t) = T_{rK}(0,t) = 1 - Q_r(0,t) = 1 - \exp\left(-\int_0^t ds \cdot h_r(s)\right)$$

where the hazard rate for the rating state $r$ is equal to (note that we used eq. [4]):

$$[7] \quad h_r(s) = \lambda_{rK}(s) = -\lambda_{rr}(s)$$

In a more general case of full-blown transition matrices, where intermediate states are possible, the instantaneous forward default rates depend on the future rating state of the issuer, and the cumulative survival probability is given by an expected value of the integrated hazard rates over all possible paths $r(s)$ of the rating transitions:

$$[8] \quad Q_r(0,t) = \sum_{q \neq K} T_{rq}(0,t) = \mathrm{E}\left\{\exp\left(-\int_0^t ds \cdot \lambda_{r(s)K}(s)\right) \middle| r(0) = r\right\}$$

where the probability density of transitioning from one rating to another at future time $s$ is governed by the generator matrix $\Lambda(s)$.

Equation [8] demonstrates how the modeling of rating transitions via a continuous-time Markov process corresponds to a fairly rich dynamic of future hazard rates, even in the case of a stationary generator matrix, where the changes in the effective future hazard rate are driven only by the random realization of the rating paths. In the case of time-dependent generator matrix models, one would obtain even more complex hazard rate dynamics, where the changes in the effective future hazard rate are also modulated by the time-dependent changes in the future rating-specific transition probabilities to a default state.

This is schematically illustrated in Figures 4 and 5. Figure 4 shows the evolution of hazard rates in a stationary transition generator model where the levels of default rates as functions of a current state are constant. Figure 5 shows the evolution of hazard rates when these levels themselves are subject to change along with the generator matrix. For simplicity, we have shown precisely the same timing of rating transitions and the ultimate default (transition to rating D). In reality, the transition timings will also be different under the two models.

**Figure 4.    Hazard rate dynamics under the stationary rating transitions model**

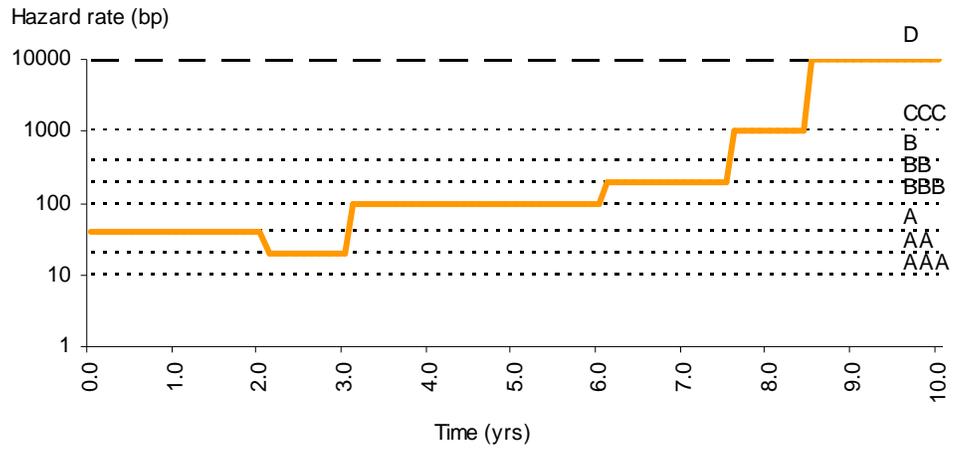

*Source: Lehman Brothers, simulation*

**Figure 5.    Hazard rate dynamics under the non-stationary rating transitions model**

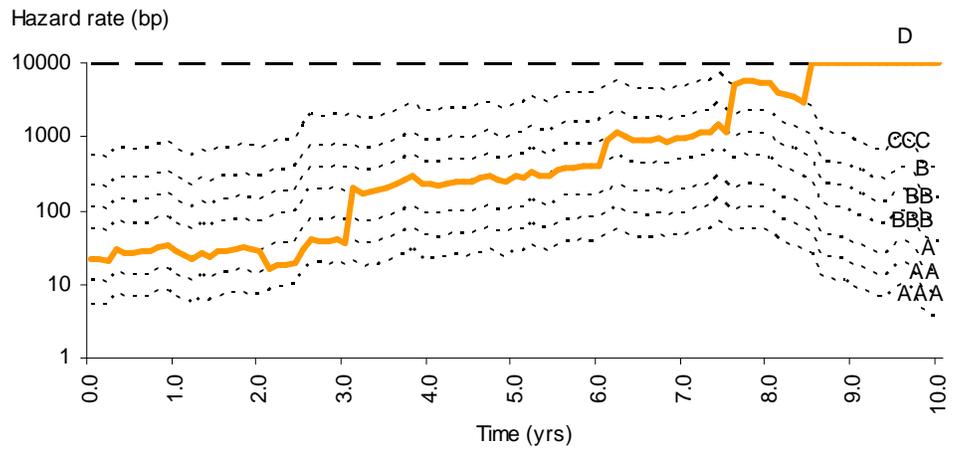

*Source: Lehman Brothers, simulation*

## ESTIMATION OF TRANSITION MATRICES

Having defined the ratings-based credit risk model in the previous section, we now turn to the estimation of the model using the historical data. First we present a brief overview of the commonly used cohort estimators that are reported by rating agencies and followed by many practitioners and explain their shortcomings. Next, following the ideas of Lando and coworkers, we will define a continuous-time estimation procedure for the generator matrix of transitions. Finally, we will extend these academic models to allow for time-weighted estimation procedures, which we believe provide a better practical solution to the often posed question "what is the current estimate of the rating transition probability?"

### Cohort Estimator for the Transition Probability Matrix

The simplest estimator of the historical transition probabilities is obtained by counting the number of times a particular type of transition $N_{ij}$ (from rating $i$ to rating $j$) has actually occurred during the observation period, and dividing it by the number of firms $N_i$ that entered the observation period with that particular initial rating $i$. This so-called cohort estimator is regularly published by major credit rating agencies such as Moody's and Standard and Poor's (see for example Carty [1993] and Carty and Fons [1997]). The name "cohort" comes from the particular implementation which tracks the transitions by the "cohort year" in which they occurred – e.g. the transition matrices for the "1990-2000 cohort":

[9]    $$T_{ij} = \frac{N_{ij}}{N_i} \text{ for } j \neq i$$

Note that if there were no transitions from rating $i$ to rating $j$ in the chosen cohort, then the estimate of the corresponding transition probability is equal to 0.

This approach has several important pitfalls:

- Rating migrations are rare events, therefore a long history is needed to get reliable estimates. This also makes it difficult to incorporate model dynamics since we cannot afford to divide the sample into smaller pieces.

- The choice of cohort time windows is somewhat artificial. Moody's convention for one-year estimates is [Jan. 1 of year $T$; Jan. 1 of year $T+1$].

- Censored data: the information about the survival time of an issuer in a given rating state both prior to the estimation window and after the end of the window is discarded. Namely, if an issuer were downgraded from A to Baa in February, we would not make use of the information about how long it had been rated A prior to the downgrade.

- Data grouping: if company X was rated A for 11 months before transition to Baa, while company Y was rated A for only 1 month, both will contribute the same to [9].

Carty and Fons (1993) suggest a remedy to the censored data problem using Weibull distributions. However, even with this modification, the cohort estimator remains susceptible to influence by the choices regarding the inclusion rules for rare transition events.

The cohort estimator is also generally inconsistent with the Markov assumption for rating transitions. The estimators for two adjacent time windows [0, u] and [u, s] and for the combined window [0, s] may not satisfy the condition [4]. While this may be taken as a signal of non-Markov behavior of ratings, we prefer to think of it as a deficiency of the estimator.

### Maximum Likelihood Estimator for the Transition Generator Matrix

Before we construct the likelihood function for the continuous-time transition generator matrix, we must first define what constitutes an observation. In this context, unlike the previous subsection, we would like to emphasize not only the event of the transition itself, but also the timeline of the rating process, starting from the date when the rating was first observed (either the date of the rating assignment or the first date of the observation period), followed by the time interval during which the rating has remained unchanged, and finishing on the date when the rating state ceased to be valid, which in turn could be due to transition to another rating, default, rating withdrawal, or the end of the observation period.

These various types of observations are illustrated in Figure 6, which mirrors the schema of the transition events table that we constructed using the Moody's default database. We show two examples of fully observed events (a ratings upgrade and a default), and three examples of censored events, where only partial information was observed – a rating withdrawal, a rating surviving until the end of the observation window, and a rating observation starting at the beginning of the observation window. The last event is called "left censored" because the censoring (incomplete information) is at the left (earlier) time boundary, while the preceding two cases are called "right censored" because the incomplete information pertains to the right (latter) time boundary.

**Figure 6.    Transition events, observation window 1/1/1990 – 10/30/2004**

| Event Type | Event # | Issuer | Start Date | Start State | End Date | End State |
|---|---|---|---|---|---|---|
| Fully Observed | 1 | A | 1/1/1992 | Baa1 | 2/15/1995 | A3 |
| Fully Observed | 2 | B | 5/15/1994 | Ba3 | 12/15/1998 | D |
| Right Censored | 3 | C | 4/10/1996 | Aa2 | 6/15/2001 | RW |
| Right Censored | 4 | D | 10/10/2002 | A1 | 10/30/2004 | A1 |
| Left Censored | 5 | E | 1/1/1990 | Ba2 | 6/30/1995 | Baa1 |

*Source: Lehman Brothers, sample*

In the continuous-time setting, each of these observations can be further decomposed into a series of mini-observations referring to consecutive short time periods, such that when strung together these mini-observations constitute the chosen line in Figure 6. For example, the observation that issuer A started in rating Baa1 on 1/1/1992 and transited to rating A3 on 2/15/1995 is reinterpreted as a mini-observation that it remained at rating Baa1 for the one-day period [1/1/1992, 1/2/1992], followed by a mini-observation that it remained at rating Baa1 for the next period [1/2/1992, 1/3/1992], and so on until the final mini-observation that the issuer transited from rating Baa1 to rating A3 during the last one-day period [2/14/1995, 2/15/1995].

We will assume in what follows a strong independence between the observations of the rating transition process. Namely, we will assume that:

- The rating transition process is Markov, i.e. all transition probabilities for each infinitesimal time period depend only on the ratings at the beginning of the period. Thus we explicitly exclude the ratings momentum and other path-dependencies.

- Rating transitions across issuers for each infinitesimal time period are conditionally independent, given their respective rating states and the rating-dependent transition probabilities at the beginning of the period.

The latter assumption is consistent with the recent empirical tests of conditional independence of the default times across issuers reported by Das, Duffie and Kapadia (2004), who show that after including the contemporaneous estimates of hazard rates and macro factors as explanatory variables, there is only very mild evidence for excess clustering of default.

Under these conditional independence assumptions, the complete likelihood function for the entire set of observations can be written as a product of likelihood functions for each event row, which in turn can be written as a product of mini-likelihood functions for each infinitesimal time period. Therefore, the total log-likelihood function is simply a sum of mini-log-likelihood functions across all observations.

Let us enumerate the observation rows by $k$, and denote the initial rating for the row by $r_k^s$, the final rating for the row by $r_k^e$. With these notations, we can write down the mini-log-likelihood functions for small time intervals as follows (here $L_t(k, r_k^s)$ corresponds to a time period when no change is observed, $L_t(k, r_k^s, r_k^e)$ corresponds to a time period when a transition was observed, and finally $L_t(k, r_k^s, C)$ corresponds to a time period when a right-censored observation occurred.

$$[10] \quad \begin{cases} L_t(k, r_k^s) & = & -\sum\limits_{q \neq r_k^s} \lambda_{r_k^s q}(t) \cdot dt \\ L_t(k, r_k^s, r_k^e) & = & \log\left(\lambda_{r_k^s r_k^e}(t) \cdot dt\right) \\ L_t(k, r_k^s, C) & = & -\sum\limits_{q \neq r_k^s} \lambda_{r_k^s q}(t) \cdot dt \end{cases}$$

Taking a sum of these mini-log-likelihood functions across the time periods corresponding to a given observation row, and them summing across rows, we obtain the following result:

$$[11] \quad L = -\sum_k \left( \int_{T_k^{start}}^{T_k^{end}} \left( \sum_{q \neq r_k^s} \lambda_{r_k^s q}(t) \right) \cdot dt \right) + \sum_k \log\left(\lambda_{r_k^s r_k^e}(t) \cdot dt\right) \cdot \left(1 - I_k^{RC}\right)$$

Here, $I_k^{RC}$ is the indicator for a right-censored event. Note that the left-censoring does not influence the log-likelihood function at all. We denoted by $T_k^{start}$ and $T_k^{end}$ the effective starting and ending time for the row $k$, respectively. Note that both the start and end times incorporate the chosen observation window, and possibly differ from the starting $t_k^{start}$ and ending $t_k^{end}$ times of a given rating transition event recorded in the database (this will be important for rolling estimates of the transition probabilities):

$$[12] \quad \begin{cases} T_k^{start} & = & \max\left[t_k^{start}, T_{window}^{start}\right] \\ T_k^{end} & = & \min\left[t_k^{end}, T_{window}^{end}\right] \end{cases}$$

With these definitions, we can rewrite the right-censored indicator in eq. [11] explicitly using indicators for a rating withdrawal and for an incomplete observation:

$$[13] \quad 1 - I_k^{RC} = \left(1 - I_{\left\{r_k^{end} = RW\right\}}\right) \cdot \left(1 - I_{\left\{t_k^{end} > T_{window}^{end}\right\}}\right)$$

Assuming that the rating transition process is stationary, i.e. $\lambda_{ij} = const$, we can easily obtain the maximum likelihood estimator of the transition generator matrix (Lando [2000], Lando and Skodeberg [2004], Christensen, Hansen and Lando [2004] and Fidelius, Lando and Nielsen [2004]):

[14] $$\hat{\lambda}_{ij} = \frac{\sum_k I_{\{r_k^{start}=i\}} \cdot I_{\{r_k^{end}=j\}} \cdot \left(1 - I_{\{r_k^{end}=RW\}}\right) \cdot \left(1 - I_{\{r_k^{end}>T_{window}^{end}\}}\right)}{\sum_k \left(T_k^{end} - T_k^{start}\right) \cdot I_{\{r_k^{start}=i\}}} \quad \text{for } i \neq j$$

In other words, the maximum likelihood estimator for the element of the generator matrix corresponding to a transition from rating $i$ to rating $j$ is equal to the count of all non-right-censored transitions within the observation window, divided by the total time that any issuer spent in the state with rating $i$ (whether the event was censored or not). Note that since the numerator includes a condition that the final rating state is $j$, then the indicator for rating withdrawal will automatically drop out (we retained it in [14] for completeness).

### Time-Weighted Maximum Likelihood Estimator

In the previous section, following Lando *et al.*, we derived the maximum likelihood estimator for the transition generator matrix assuming that it is constant over time. However, as mentioned earlier, the empirical studies as well as the fundamental understanding of the credit markets suggest that one must account for time variation of the hazard rates in order to explain the observed patterns of rating transitions. Without allowing for such variation, there is a significant clustering of downgrades and defaults in times of market downturns, such as 1990-1992 and 2000-2002 (see Das, Duffie and Kapadia for detailed discussion of default risk clustering in particular).

Indeed, equation [14] implies that a company that was downgraded from Baa to Ba in 1996 having survived five years at Baa, is treated similarly to a company that was downgraded in 2002 having survived only a year at Baa. In doing so, the typically longer survival time of the companies during the mid-1990s would significantly increase the denominator of [14] and depress the estimate of the corresponding downgrade probability.

For many practitioners it is obvious that the benign credit risk environment in 1996 had little in common with the severe credit downturn during 2002. The shorter lifespan of a company in a given rating category during 2002 was not an exception but a rule. Hence, if we were trying to estimate the rating transition risk in 2002, it might not have been meaningful to dilute the information contained in the recent volatile history of transitions by adding to it a large number of much more tempered transition events from a long time ago.

The cohort estimator approach deals with this problem by imposing an observation window, such as the trailing three-year or a trailing five-year window. We already noted some of the drawbacks associated with the imposition of such windows. One additional comment is that sharp windows pose an even bigger problem when the observation falls off the back end of the window than when it enters from the front. Indeed, while the potentially large impact of a new observation may seem reasonable, an equally large impact from the fact that an old observation was, say 5.2 years ago compared with 4.9 years ago, in the case of a sharp 5-year window does not seem justified. It would be much better if the influence of all old observations was gradually fading away, never quite dropping entirely out of consideration and never changing too much from one month to another as we move forward in time.

One of the simplest ways to account for time variation of transition intensities and to solve the problem of observation window selection is to apply a time-weighting scheme to the maximum likelihood estimator. The most common such scheme is the geometric (or exponential) time-decay with half-life $T_H$, which assigns the following weight to an event that occurred at time $t$ being observed at later time $T$, e.g. the rolling estimation date:

[15]     $$w(t,T) = 2^{-\frac{T-t}{T_H}} = \exp\left(-\frac{T-t}{T_H^{\exp}}\right) \qquad \text{where} \quad T_H^{\exp} = \frac{T_H}{\log(2)}$$

The weight assigned to an event that occurred at time $T_H$ prior to observation date is 2 times less than for an even that occurred just an instant ago.

To estimate a weighted log-likelihood function, we remember that it was constructed from independent mini-likelihood functions [10]. Assigning each mini-observation a weight corresponding to the time of the mini-event is the same as taking each of the mini-likelihood functions to the power of the corresponding weight, or else equivalently multiplying each mini-log-likelihood function by the chosen weight. Therefore, we obtain:

[16]     $$\begin{aligned} L(T) \;=\; & -\sum_k \left( \int_{T_k^{start}}^{T_k^{end}} w(t,T) \cdot \left( \sum_{q \neq r_k^s} \lambda_{r_k^s q}(T) \right) \cdot dt \right) \\ & + \sum_k w(T_k^{end}, T) \cdot \log\left( \lambda_{r_k^s r_k^e}(T) \cdot dt \right) \cdot \left( 1 - I_k^{RC} \right) \end{aligned}$$

Note that we explicitly acknowledged the dependence of the transition generator estimate on the rolling observation date – as the estimation time $T = T_{window}^{end}$ goes on the maximum likelihood estimator will change both because of the new events being observed and because of the changing weights. The time-weighted analog of equation [14] is given by:

[17]     $$\hat{\lambda}_{ij}(T) = \frac{\displaystyle\sum_k w\left(T_k^{end}, T\right) \cdot I_{\{r_k^{start}=i\}} \cdot I_{\{r_k^{end}=j\}} \cdot \left(1 - I_{\{t_k^{end}>T\}}\right)}{\displaystyle\sum_k I_{\{r_k^{start}=i\}} \cdot \int_{T_k^{start}}^{T_k^{end}} w(t,T) \cdot dt}$$

Substituting the specific form of the exponentially-decaying weights, we obtain:

[18]     $$\hat{\lambda}_{ij}(T) = \frac{\displaystyle\sum_k 2^{-\frac{T-T_k^{end}}{T_H}} \cdot I_{\{r_k^{start}=i\}} \cdot I_{\{r_k^{end}=j\}} \cdot \left(1 - I_{\{t_k^{end}>T\}}\right)}{\displaystyle T_H \cdot \sum_k I_{\{r_k^{start}=i\}} \cdot \left( 2^{-\frac{T-T_k^{end}}{T_H}} - 2^{-\frac{T-T_k^{start}}{T_H}} \right)}$$

We can see that in the limit $T_H \to \infty$ the estimate converges to the unweighted case [14]. We can also approximate the cohort estimate [9] in a special limiting case where the half-life parameter is much greater than the time to observed transitions, but still much smaller than all survival periods $T - T_k^{end} << T_H << T_k^{end} - T_k^{start}$ for all $k$. Since this limit is very unlikely to be satisfied, we conclude that the cohort estimator is indeed not optimal.

## A Simple Example

To illustrate both the cohort and the continuous-time estimation procedure, consider a simple example, which is a modified version of the one suggested by Lando and Skodeberg (2004).

Let us assume a rating system with two non-default rating categories A, B and a default category D. Assume that we observe over one year the history of 20 firms, of which 10 start in category A and 10 in category B. Assume that over the year of observation, one A-rated firm changes its rating to category B after three months and stays there for the rest of the year. Assume that over the same period, one B-rated firm is upgraded after nine months and remains in A for the rest of the period, and a firm which started in B defaults after six months and stays there for the remaining part of the period (see Figure 7, where we have highlighted the rows with non-trivial transition activity. Dates are measured in months).

**Figure 7.    Example transition events table, observation period [0, 12] months**

| Event # | Issuer | Start Date | Start State | End Date | End State |
|---|---|---|---|---|---|
| 1 | 1 | 0 | A | 3 | B |
| 2 | 1 | 3 | B | 12 | B |
| 3 | 2 | 0 | A | 12 | A |
| 4 | 3 | 0 | A | 12 | A |
| 5 | 4 | 0 | A | 12 | A |
| 6 | 5 | 0 | A | 12 | A |
| 7 | 6 | 0 | A | 12 | A |
| 8 | 7 | 0 | A | 12 | A |
| 9 | 8 | 0 | A | 12 | A |
| 10 | 9 | 0 | A | 12 | B |
| 11 | 10 | 0 | A | 12 | B |
| 12 | 11 | 0 | B | 9 | A |
| 13 | 11 | 9 | A | 12 | A |
| 14 | 12 | 0 | B | 6 | D |
| 15 | 13 | 0 | B | 12 | B |
| 16 | 14 | 0 | B | 12 | B |
| 17 | 15 | 0 | B | 12 | B |
| 18 | 16 | 0 | B | 12 | B |
| 19 | 17 | 0 | B | 12 | B |
| 20 | 18 | 0 | B | 12 | B |
| 21 | 19 | 0 | B | 12 | B |
| 22 | 20 | 0 | B | 12 | B |

*Source: Lehman Brothers, simulated example*

By straightforward application of the formulas [9], [14] and [18] we can now calculate the cohort estimator of the 1-year transition probability matrix, and the maximum likelihood estimator of the 1-year generator matrix under a stationary assumption, and under a time-weighted assumption with a half-life of six months. Having obtained the estimates of the generator matrices, we can also calculate, via a matrix exponential, the estimates of the 1-year transition probability matrices for the latter two cases. The results of this exercise are presented in Figure 8. We show the estimates for the generator matrices on the left and the corresponding estimates for the transition probability matrices on the right. For the cohort method, the generator matrix is calculated from the transition matrix by taking a matrix log.

**Figure 8.    Estimates for transition matrices under various assumptions**

| | Transition Generator Matrix | | | | Transition Probability Matrix | | |
|---|---|---|---|---|---|---|---|
| | | | | | **Cohort Estimator** | | |
| | **A** | **B** | **D** | ⇐ mlog | **A** | **B** | **D** |
| **A** | -0.1121 | 0.1183 | -0.0063 | | 0.9000 | 0.1000 | 0 |
| **B** | 0.1183 | -0.2304 | 0.1121 | | 0.1000 | 0.8000 | 0.1000 |
| **D** | 0 | 0 | 0 | | 0 | 0 | 1 |

| | **Stationary Continuous-Time Estimator** | | | | | | |
|---|---|---|---|---|---|---|---|
| | **A** | **B** | **D** | mexp ⇒ | **A** | **B** | **D** |
| **A** | -0.3158 | 0.3158 | 0.0000 | | 0.7412 | 0.2454 | 0.0134 |
| **B** | 0.1000 | -0.2000 | 0.1000 | | 0.0777 | 0.8312 | 0.0911 |
| **D** | 0 | 0 | 0 | | 0 | 0 | 1 |

| | **Time-Weighted Continuous-Time Estimator** | | | | | | |
|---|---|---|---|---|---|---|---|
| | **A** | **B** | **D** | mexp ⇒ | **A** | **B** | **D** |
| **A** | -0.4566 | 0.4566 | 0.0000 | | 0.6544 | 0.3283 | 0.0173 |
| **B** | 0.1333 | -0.2276 | 0.0943 | | 0.0959 | 0.8190 | 0.0851 |
| **D** | 0 | 0 | 0 | | 0 | 0 | 1 |

*Source: Lehman Brothers*

There are several observations that can be made by comparing the results in Figure 8. First of all, we note that even in this simple example it becomes apparent that the cohort estimator is not fully consistent with a continuous-time Markov description of the rating transition process. Indeed, the highlighted cells in the A-D transition rows show that the cohort estimate of direct 1-year default probability for category A being exactly zero could only be reconciled with the continuous-time framework if the transition hazard rate from A to D were *negative*.

We can understand this better by looking at the case of the stationary continuous-time estimate. Here, the transition hazard rate from A to D actually is zero. However, it still leads to a non-zero 1-year transition probability from A to D, due to the possibility of multi-stage transitions like A-B-D, A-B-A-B-D, etc. The only elements of the 1-year transition matrix that are exactly zero correspond to forbidden transitions from D to A or B. Consequently, it is impossible to find a generator matrix with non-negative transition probabilities that would produce a zero probability for the 1-year transition A-D.

Finally, comparing the stationary and the time-weighted continuous-time generator matrices and their corresponding 1-year transition probabilities, we can see that the impact of the weighting scheme can be quite substantial. Moreover, the way it affects the estimates is driven by a complex interplay between the time of the transition event (more recent transitions increase the corresponding hazard rate) and the duration of the stay in the initial rating (longer survival times decrease the corresponding hazard rate). For example, the A-B transition probability went up by a much larger factor than the B-D transition probability went down, even though both are driven by a single event, because the survival time length (three months vs. six months) had a bigger impact on the hazard rate estimates than the event timing (nine months ago vs. six months ago) for these two transition events.

## Parametric Smoothing of the Transition Generator Matrix

The main difficulty in the empirical estimation of the rating transition process is the rarity of the observed events, coupled with a relatively large number of the model parameters that need estimating. For a notched rating transition generator matrix (with rating states Aaa, Aa1, Aa2, …, B3, Caa, D) we have 17x18 non-zero entries with 17 constraints, requiring us to estimate as many as 289 parameters. This is a lot by any measure, even if the events were not so rare. To reliably estimate so many parameters, some of which are expected to be very small numbers, one would need perhaps 100 times as many (or more) transition events, which is roughly the same order of magnitude as the entire size of Moody's transition database. Since we are interested primarily in the "dynamic" estimates of transition probabilities over relatively short time horizons using the exponentially weighted approach introduced earlier, we would likely face a severe problem with the number of "effective" observations in the sample being much less than required for the desired accuracy.

Even for a much more parsimonious letter-grade matrix (with rating states Aaa, Aa, A, Baa, Ba, B, Caa, D) we have 7x8 non-zero entries, with 7 constraints, leaving us with 49 model parameters. The reduction in the model complexity is deceptive, however, since by turning to a coarse-grained picture of rating transitions we simultaneously lose the lion's share of the observations – for example, we can no longer derive much from a transition from Baa1 to Baa3 and must wait until the corresponding issuer gets downgraded to the next letter level until it contributes to the (observed event component of the) likelihood function.

The natural solution for this problem is to impose more structure on the model and to reduce the number of independent model parameters, so that fewer observations would be sufficient to reliably estimate them. Some of the commonly cited guiding principles for doing this are:

- The transition probabilities must be monotonic, i.e. a transition to a farther state must be less probable than a transition to a nearer state.

- An exception is made for default probabilities, since the default is an absorbing state and accumulates the probabilities over time. However, this exception is not necessary when considering continuous-time transition generating matrices.

- The shape of the monotonically decreasing transition probabilities for upgrades and downgrades need not be the same, e.g. a one-notch upgrade need not have the same probability as a one-notch downgrade.

Arguably, the best way to define a meaningful parameterization is by using an underlying model that "explains" the rating transitions with fewer model parameters. The well known CreditMetrics model (see Gupton, Finger and Bhatia [1997]) uses an underlying Merton-like structural model as a guiding tool for smoothing the transition probability matrix. They also discuss a variety of purely statistical fitting techniques. A completely different perspective is taken by Keenan and Sobehart (2002), who derive the parameterization of the transition probability matrix from a behavioral model of analyst decisions. Both of these studies are focused on the finite-horizon transition probability matrices, and the resulting smooth estimates are not guaranteed to correspond to a well-defined transition generating matrix, although the discrepancies are usually quite small.

We would like to propose yet another parameterization that is particularly well suited to the continuous-time estimation framework. It is motivated by an observation made earlier that agency rating actions often lag the news about the issuer's financial health. This lag, which is partly maintained by the agencies to ensure that only the reliable and material changes to a company's long-term creditworthiness get reflected in the rating upgrades or downgrades, has a side-effect that sometimes the required action is unusually strong. Indeed, one could argue that if the rating analysts had kept up with every bit of news and market sentiment they would perhaps change the ratings much more frequently but only by the smallest amount possible, i.e. by one notch. Such behavior would be, in a sense, in the spirit of the market-driven credit risk assessment advocated by Moody's KMV (Kealhofer [2003]).

We can adopt a slightly less demanding view of the rating assignment process by assuming that there is an underlying unobservable continuous-time "shadow" rating process which can only proceed by single-notch upgrades or downgrades, but that the outcomes of this shadow process are revealed at random lag times, which are exponentially distributed with some characteristic time scale $\theta$. As a result, there will be some probability that by the time the outcome is observed the rating has actually drifted by more than a single notch. Because multiple-notch upgrades and downgrades in this model correspond to compounded probabilities of single-notch rating changes (which are presumably small to begin with), and the bigger the rating change the more compounding is required, then we might expect that the model would naturally predict the desired monotonic behavior for transition probabilities.

The formal definition of the transition generator matrix in our parametric model is given by:

$$[19] \quad \Lambda = \theta \cdot \int_0^{+\infty} d\tau \cdot \exp\left(-\frac{\tau}{\theta}\right) \cdot \operatorname{mexp}(G \cdot \tau) - I = (I - \theta \cdot G)^{-1} - I$$

Here, $G$ is the generator matrix for the "shadow" rating process which, by assumption, has a tri-diagonal form allowing only nearest neighbor rating transitions:

$$[20] \quad G = \begin{pmatrix} -g_{12} & g_{12} & 0 & \cdots & 0 \\ g_{21} & -(g_{21}+g_{23}) & g_{23} & 0 & \vdots \\ \vdots & \vdots & \vdots & \vdots & \vdots \\ \vdots & 0 & g_{K-1,K-2} & -(g_{K-1,K-2}+g_{K-1,K}) & g_{K-1,K} \\ 0 & 0 & \cdots & 0 & 0 \end{pmatrix}$$

The integral in our definition corresponds to summing over all possible "revelation" times, which are exponentially distributed with characteristic time scale $\theta$. The matrix exponent under the integral provides the outcome probabilities for the compounded single-notch transition processes. Since the matrix exponents give manifestly valid transition probabilities, and since we are averaging them with positive weights that sum up to 1, then the integral in equation [19] has the properties of a nice transition probability matrix. Correspondingly, by subtracting an identity matrix from it, we obtain a generator matrix that is guaranteed to have positive off-diagonal elements and rows that sum up to zero, and is therefore a valid transition generator matrix. In a sense, the matrix $G$ acts as a generator matrix for the generator matrix $\Lambda$ (we call it a "gengen" matrix).

The integral itself is easy to take, and results in a simple matrix inverse, as shown on the right-hand side of eq. [19]. As expected, in the limit of very small characteristic times $\theta \to 0$ while holding $G$ constant, the rating transition generator matrix becomes tri-diagonal as well, being simply proportional to the gengen matrix $\Lambda \to \theta \cdot G$.

With the parameterization defined by eqs. [19] and [20], we have only 2*(K-1) variables for a K-state rating system (one variable is the characteristic time, and the other variables are the K-1 elements of the upper diagonal and K-2 elements of the lower diagonal). There is, however, a residual scaling freedom that needs to be fixed, namely the simultaneous change of $\theta \rightarrow \alpha \cdot \theta$ and $G \rightarrow \alpha^{-1} \cdot G$ leaves the transition generator matrix $\Lambda$ unchanged. Therefore, we can simply arbitrarily fix the value $\theta = 1$ and only consider the off-diagonal elements of the gengen matrix to be independent. Thus, for a full 18-state notched rating system we have to estimate only 2*K - 3 = 33 parameters as opposed to 289 values in the unconstrained generator matrix. With this few parameters we can indeed hope to have sufficient number of observations even for relatively short estimation horizons such as one, three or five years. Of course, for any number of model parameters, the shorter horizons correspond to fewer "effective" observations and less accurate estimates, and one must balance this against the benefits derived from capturing the time-varying dynamics.

Even though we derived the parametric form of the generator matrix [19] by assuming an existence of the unobservable "shadow" rating process, we still interpret the observed rating transitions in accordance with the continuous-time inhomogeneous Markov process governed by eqs. [5] and [6]. Therefore, the time-weighted log-likelihood function is still given by the same expression [16] as before, except the elements of the transition generator matrix $\lambda_{ij}$ are now treated as explicit functions of the gengen matrix $G$ (remember that we fixed the time scale $\theta = 1$), which in turn is a function of the estimation time $T = T_{window}^{end}$. Substituting the standard constraint [4] and the exponentially decaying weights [15] into eq. [16] we obtain:

[21]
$$
\begin{aligned}
L(T) \;=\; & T_H \cdot \sum_k \left( 2^{-\frac{T - T_k^{end}}{T_H}} - 2^{-\frac{T - T_k^{start}}{T_H}} \right) \cdot \lambda_{r_k^s r_k^e}\left(G(T)\right) \\
& + \sum_k 2^{-\frac{T - T_k^{end}}{T_H}} \cdot \log\!\left(\lambda_{r_k^s r_k^e}\left(G(T)\right) \cdot dt\right) \cdot \left(1 - I_{\{r_k^{end} = RW\}}\right) \cdot \left(1 - I_{\{r_k^{end} > T\}}\right)
\end{aligned}
$$

Since the dependence of $\lambda_{ij}$ on the model parameters $g_{ij}$ is non-linear [19], we are no longer able to derive an explicit solution for the maximum likelihood estimators like eq. [18]. Instead, one has to perform the maximization numerically. However, this complication is a small price to pay for the increased robustness and accuracy of the model.

## THE DYNAMICS OF RATING TRANSITIONS

Let us now turn to the historical data. We have implemented the maximum likelihood estimation methodology with parametric smoothing described in the previous section (with adjustments for partial censoring of data discussed in the Appendix A) for a set of half-life parameters including 1, 3, 5, 10, and 20 years.

We treat the 20-year half-life estimate as the long-term transition probabilities that are unconditional on the business cycle. The 10-year half-life corresponds to an average over a single business cycle, which in recent history lasted roughly ten years.

The shorter half-lives of 1, 3, and 5 years correspond to conditional estimates and are much more sensitive to the point in the business cycle at which the estimate is made. We feel that a 1-year half-life estimate is only marginally useful since the effective number of transition events is very low and the accuracy of the estimate is poor – we produce it for completeness and for comparison with often cited 1-year cohort estimates. However, the 3- and 5-year half-life estimates are actually quite robust and reflect a meaningful difference in conditioning information in comparison with long-term estimates as well as in comparison with each other.

There are many uses of such transition matrix estimates. Depending on the context, either the longer-term, unconditional matrices, or the shorter-term, conditional matrices may be appropriate. For example, if an investor is interested in the analysis of credit losses for a hypothetical 10-year CLO backed by a diversified pool of loans about which she knows only their ratings distribution, then applying the through-the-cycle average estimate such as the 10-year half-life generator matrix and deriving the corresponding cumulative default probabilities for a 10-year horizon would seem reasonable. On the other hand, if an asset manager is inquiring about the upgrade-to-downgrade ratio in the next year, then using an estimate based on a short 3-year half-life is more likely to yield a useful answer (again, we caution against a 1-year half-life because of less accuracy of the estimates).

In this section we discuss the dynamics of the estimated transition probabilities obtained from our model, and highlight the degree to which it can impact the frequently used values such as long-horizon default probabilities. We then briefly consider some stylized facts about the predictability of rating transitions.

### Rolling Estimates of Historical Transition Probabilities

The simplest and most common use of the credit rating transition models is to obtain various cumulative probabilities for a given time horizon such as 1, 5 or 10 years, under the assumption that the present estimate of the transition matrix (or the generator matrix in continuous-time formalism) will remain intact. However, this does not prevent one changing the present estimate as time goes on and new observations accrue in the database. In this subsection we show the results of such rolling estimates and discuss their dynamics.

We begin with the estimation of the long-term unconditional transition matrices, i.e. those estimated with a half-life of 20 years, as of 1/1/2005. Figure 9a shows the estimate of the gengen matrix, 9b shows the corresponding transition generator matrix, 9c shows the 1-year transition probability matrix and 9d shows the 10-year transition probability matrix.

For comparison, we show a similar set of estimates as of 1/1/2003 with a 3-year half-life in Figures 10a-d. As we can see, the greater downgrade risk of the period 2000-2003 has been captured in these estimates. While the differences for the generator matrices may seem small, one can see from the comparison of the transition probabilities (Figures 9c and 10c), and especially from comparison of the 10-year transition probability matrices (Figures 9d and 10d), that estimates for both downgrade and default risk in 2003 were markedly higher than the corresponding unconditional ones.

In Figure 11a we show the often cited Moody's idealized cumulative default probabilities, and widely used weighted average rating factor (WARF) values for each rating which are numerically equal to 10-year cumulative default probability measured in basis points. For comparison, in Figure 11b we show the same numbers obtained from our unconditional long-term model. As we can see, our estimates are higher (see the WARF column for comparison). Figure 11c demonstrates the same calculation for a short-term estimate as of 1/1/2003, and we see that the 10-year cumulative default estimates in our model at that time would have been several times higher than Moody's WARF numbers would suggest. Even for shorter horizons such as 5 or 3 years, the differences are quite dramatic. This is, of course, not surprising – market practitioners were well aware of the much higher credit risk during the early 2000s. However, it is notable that our model correctly captures this effect.

**Figure 9a.  Gengen matrix estimated as of 1/1/2005 with 20-year half-life**

| Rating | AAA | AA1 | AA2 | AA3 | A1 | A2 | A3 | BAA1 | BAA2 | BAA3 | BA1 | BA2 | BA3 | B1 | B2 | B3 | CCC | D |
|---|---|---|---|---|---|---|---|---|---|---|---|---|---|---|---|---|---|---|
| AAA | (0.1442) | 0.1442 | - | - | - | - | - | - | - | - | - | - | - | - | - | - | - | - |
| AA1 | 0.1403 | (0.6206) | 0.4803 | - | - | - | - | - | - | - | - | - | - | - | - | - | - | - |
| AA2 | - | 0.2363 | (0.7020) | 0.4656 | - | - | - | - | - | - | - | - | - | - | - | - | - | - |
| AA3 | - | - | 0.2310 | (0.6505) | 0.4195 | - | - | - | - | - | - | - | - | - | - | - | - | - |
| A1 | - | - | - | 0.2716 | (0.6889) | 0.4173 | - | - | - | - | - | - | - | - | - | - | - | - |
| A2 | - | - | - | - | 0.3083 | (0.7991) | 0.4907 | - | - | - | - | - | - | - | - | - | - | - |
| A3 | - | - | - | - | - | 0.4460 | (1.0056) | 0.5595 | - | - | - | - | - | - | - | - | - | - |
| BAA1 | - | - | - | - | - | - | 0.5932 | (1.3488) | 0.7556 | - | - | - | - | - | - | - | - | - |
| BAA2 | - | - | - | - | - | - | - | 0.5820 | (1.2396) | 0.6576 | - | - | - | - | - | - | - | - |
| BAA3 | - | - | - | - | - | - | - | - | 0.6713 | (1.4165) | 0.7453 | - | - | - | - | - | - | - |
| BA1 | - | - | - | - | - | - | - | - | - | 0.8587 | (1.9216) | 1.0629 | - | - | - | - | - | - |
| BA2 | - | - | - | - | - | - | - | - | - | - | 1.0668 | (2.5037) | 1.4369 | - | - | - | - | - |
| BA3 | - | - | - | - | - | - | - | - | - | - | - | 0.8784 | (1.8644) | 0.9860 | - | - | - | - |
| B1 | - | - | - | - | - | - | - | - | - | - | - | - | 0.7535 | (2.2282) | 1.4747 | - | - | - |
| B2 | - | - | - | - | - | - | - | - | - | - | - | - | - | 1.1658 | (2.6698) | 1.5040 | - | - |
| B3 | - | - | - | - | - | - | - | - | - | - | - | - | - | - | 0.8520 | (2.0303) | 1.1783 | - |
| CCC | - | - | - | - | - | - | - | - | - | - | - | - | - | - | - | 0.2702 | (0.7325) | 0.4623 |
| D | - | - | - | - | - | - | - | - | - | - | - | - | - | - | - | - | - | - |

**Figure 9b.  Transition generator matrix as of 1/1/2005 with 20-year half-life**

| Rating | AAA | AA1 | AA2 | AA3 | A1 | A2 | A3 | BAA1 | BAA2 | BAA3 | BA1 | BA2 | BA3 | B1 | B2 | B3 | CCC | D |
|---|---|---|---|---|---|---|---|---|---|---|---|---|---|---|---|---|---|---|
| AAA | (0.1159) | 0.0822 | 0.0242 | 0.0071 | 0.0019 | 0.0005 | 0.0001 | 0.0000 | 0.0000 | 0.0000 | 0.0000 | 0.0000 | 0.0000 | 0.0000 | 0.0000 | 0.0000 | 0.0000 | 0.0000 |
| AA1 | 0.0800 | (0.3479) | 0.1917 | 0.0565 | 0.0147 | 0.0036 | 0.0010 | 0.0003 | 0.0001 | 0.0000 | 0.0000 | 0.0000 | 0.0000 | 0.0000 | 0.0000 | 0.0000 | 0.0000 | 0.0000 |
| AA2 | 0.0116 | 0.0943 | (0.3602) | 0.1886 | 0.0491 | 0.0122 | 0.0032 | 0.0008 | 0.0003 | 0.0001 | 0.0000 | 0.0000 | 0.0000 | 0.0000 | 0.0000 | 0.0000 | 0.0000 | 0.0000 |
| AA3 | 0.0017 | 0.0138 | 0.0936 | (0.3395) | 0.1719 | 0.0427 | 0.0113 | 0.0030 | 0.0011 | 0.0003 | 0.0001 | 0.0000 | 0.0000 | 0.0000 | 0.0000 | 0.0000 | 0.0000 | 0.0000 |
| A1 | 0.0003 | 0.0023 | 0.0158 | 0.1113 | (0.3508) | 0.1612 | 0.0428 | 0.0112 | 0.0042 | 0.0013 | 0.0004 | 0.0001 | 0.0001 | 0.0000 | 0.0000 | 0.0000 | 0.0000 | 0.0000 |
| A2 | 0.0001 | 0.0004 | 0.0029 | 0.0204 | 0.1191 | (0.3755) | 0.1657 | 0.0435 | 0.0161 | 0.0049 | 0.0014 | 0.0005 | 0.0003 | 0.0001 | 0.0000 | 0.0000 | 0.0000 | 0.0000 |
| A3 | 0.0000 | 0.0001 | 0.0007 | 0.0049 | 0.0287 | 0.1506 | (0.4195) | 0.1523 | 0.0565 | 0.0172 | 0.0050 | 0.0018 | 0.0010 | 0.0004 | 0.0002 | 0.0001 | 0.0001 | 0.0000 |
| BAA1 | 0.0000 | 0.0000 | 0.0002 | 0.0014 | 0.0080 | 0.0419 | 0.1615 | (0.4888) | 0.1898 | 0.0576 | 0.0169 | 0.0059 | 0.0033 | 0.0012 | 0.0006 | 0.0003 | 0.0002 | 0.0001 |
| BAA2 | 0.0000 | 0.0000 | 0.0001 | 0.0004 | 0.0023 | 0.0120 | 0.0462 | 0.1462 | (0.4545) | 0.1657 | 0.0485 | 0.0171 | 0.0095 | 0.0035 | 0.0016 | 0.0008 | 0.0006 | 0.0003 |
| BAA3 | 0.0000 | 0.0000 | 0.0000 | 0.0001 | 0.0007 | 0.0037 | 0.0143 | 0.0453 | 0.1691 | (0.4868) | 0.1503 | 0.0529 | 0.0294 | 0.0108 | 0.0049 | 0.0026 | 0.0018 | 0.0008 |
| BA1 | 0.0000 | 0.0000 | 0.0000 | 0.0000 | 0.0002 | 0.0013 | 0.0048 | 0.0153 | 0.0570 | 0.1731 | (0.5565) | 0.1563 | 0.0868 | 0.0318 | 0.0145 | 0.0077 | 0.0052 | 0.0024 |
| BA2 | 0.0000 | 0.0000 | 0.0000 | 0.0000 | 0.0001 | 0.0004 | 0.0017 | 0.0054 | 0.0202 | 0.0612 | 0.1569 | (0.6132) | 0.2147 | 0.0786 | 0.0360 | 0.0190 | 0.0129 | 0.0060 |
| BA3 | 0.0000 | 0.0000 | 0.0000 | 0.0000 | 0.0000 | 0.0002 | 0.0006 | 0.0018 | 0.0068 | 0.0208 | 0.0532 | 0.1313 | (0.5408) | 0.1680 | 0.0770 | 0.0407 | 0.0277 | 0.0128 |
| B1 | 0.0000 | 0.0000 | 0.0000 | 0.0000 | 0.0000 | 0.0000 | 0.0002 | 0.0005 | 0.0019 | 0.0058 | 0.0149 | 0.0367 | 0.1284 | (0.5818) | 0.1915 | 0.1012 | 0.0688 | 0.0318 |
| B2 | 0.0000 | 0.0000 | 0.0000 | 0.0000 | 0.0000 | 0.0000 | 0.0001 | 0.0002 | 0.0007 | 0.0021 | 0.0054 | 0.0133 | 0.0465 | 0.1514 | (0.6201) | 0.2008 | 0.1365 | 0.0631 |
| B3 | 0.0000 | 0.0000 | 0.0000 | 0.0000 | 0.0000 | 0.0000 | 0.0000 | 0.0001 | 0.0002 | 0.0006 | 0.0016 | 0.0040 | 0.0139 | 0.0453 | 0.1137 | (0.5886) | 0.2798 | 0.1294 |
| CCC | 0.0000 | 0.0000 | 0.0000 | 0.0000 | 0.0000 | 0.0000 | 0.0000 | 0.0000 | 0.0000 | 0.0001 | 0.0003 | 0.0006 | 0.0022 | 0.0071 | 0.0177 | 0.0642 | (0.3792) | 0.2870 |
| D | - | - | - | - | - | - | - | - | - | - | - | - | - | - | - | - | - | - |

**Figure 9c.  1-year transition probability matrix as of 1/1/2005 with 20-year half-life**

| Rating | AAA | AA1 | AA2 | AA3 | A1 | A2 | A3 | BAA1 | BAA2 | BAA3 | BA1 | BA2 | BA3 | B1 | B2 | B3 | CCC | D |
|---|---|---|---|---|---|---|---|---|---|---|---|---|---|---|---|---|---|---|
| AAA | 0.8934 | 0.0665 | 0.0256 | 0.0098 | 0.0032 | 0.0010 | 0.0003 | 0.0001 | 0.0000 | 0.0000 | 0.0000 | 0.0000 | 0.0000 | 0.0000 | 0.0000 | 0.0000 | 0.0000 | 0.0000 |
| AA1 | 0.0647 | 0.7156 | 0.1383 | 0.0543 | 0.0184 | 0.0058 | 0.0019 | 0.0006 | 0.0003 | 0.0001 | 0.0000 | 0.0000 | 0.0000 | 0.0000 | 0.0000 | 0.0000 | 0.0000 | 0.0000 |
| AA2 | 0.0122 | 0.0681 | 0.7108 | 0.1382 | 0.0476 | 0.0153 | 0.0050 | 0.0016 | 0.0007 | 0.0003 | 0.0001 | 0.0000 | 0.0000 | 0.0000 | 0.0000 | 0.0000 | 0.0000 | 0.0000 |
| AA3 | 0.0023 | 0.0133 | 0.0686 | 0.7263 | 0.1267 | 0.0414 | 0.0139 | 0.0044 | 0.0020 | 0.0007 | 0.0002 | 0.0001 | 0.0001 | 0.0000 | 0.0000 | 0.0000 | 0.0000 | 0.0000 |
| A1 | 0.0005 | 0.0029 | 0.0153 | 0.0820 | 0.7189 | 0.1177 | 0.0402 | 0.0130 | 0.0059 | 0.0022 | 0.0007 | 0.0003 | 0.0002 | 0.0001 | 0.0000 | 0.0000 | 0.0000 | 0.0000 |
| A2 | 0.0001 | 0.0007 | 0.0036 | 0.0198 | 0.0870 | 0.7038 | 0.1174 | 0.0387 | 0.0179 | 0.0066 | 0.0023 | 0.0009 | 0.0006 | 0.0003 | 0.0001 | 0.0001 | 0.0001 | 0.0000 |
| A3 | 0.0000 | 0.0002 | 0.0011 | 0.0060 | 0.0270 | 0.1067 | 0.6759 | 0.1036 | 0.0488 | 0.0183 | 0.0064 | 0.0026 | 0.0017 | 0.0007 | 0.0004 | 0.0002 | 0.0002 | 0.0001 |
| BAA1 | 0.0000 | 0.0001 | 0.0004 | 0.0020 | 0.0093 | 0.0373 | 0.1098 | 0.6323 | 0.1272 | 0.0484 | 0.0171 | 0.0071 | 0.0046 | 0.0020 | 0.0010 | 0.0006 | 0.0005 | 0.0003 |
| BAA2 | 0.0000 | 0.0000 | 0.0001 | 0.0007 | 0.0033 | 0.0133 | 0.0399 | 0.0979 | 0.6554 | 0.1114 | 0.0398 | 0.0166 | 0.0109 | 0.0047 | 0.0025 | 0.0015 | 0.0012 | 0.0008 |
| BAA3 | 0.0000 | 0.0000 | 0.0000 | 0.0003 | 0.0012 | 0.0050 | 0.0153 | 0.0381 | 0.1137 | 0.6345 | 0.0967 | 0.0408 | 0.0272 | 0.0119 | 0.0063 | 0.0038 | 0.0032 | 0.0021 |
| BA1 | 0.0000 | 0.0000 | 0.0000 | 0.0001 | 0.0005 | 0.0020 | 0.0061 | 0.0155 | 0.0468 | 0.1114 | 0.5915 | 0.0956 | 0.0645 | 0.0285 | 0.0151 | 0.0093 | 0.0078 | 0.0053 |
| BA2 | 0.0000 | 0.0000 | 0.0000 | 0.0000 | 0.0002 | 0.0008 | 0.0025 | 0.0064 | 0.0196 | 0.0471 | 0.0959 | 0.5595 | 0.1309 | 0.0586 | 0.0313 | 0.0194 | 0.0165 | 0.0112 |
| BA3 | 0.0000 | 0.0000 | 0.0000 | 0.0000 | 0.0001 | 0.0003 | 0.0010 | 0.0026 | 0.0079 | 0.0192 | 0.0395 | 0.0800 | 0.6006 | 0.1053 | 0.0567 | 0.0354 | 0.0304 | 0.0210 |
| B1 | 0.0000 | 0.0000 | 0.0000 | 0.0000 | 0.0000 | 0.0001 | 0.0003 | 0.0008 | 0.0026 | 0.0064 | 0.0134 | 0.0274 | 0.0805 | 0.5767 | 0.1138 | 0.0717 | 0.0625 | 0.0439 |
| B2 | 0.0000 | 0.0000 | 0.0000 | 0.0000 | 0.0000 | 0.0000 | 0.0001 | 0.0003 | 0.0011 | 0.0027 | 0.0056 | 0.0116 | 0.0343 | 0.0900 | 0.5550 | 0.1191 | 0.1051 | 0.0752 |
| B3 | 0.0000 | 0.0000 | 0.0000 | 0.0000 | 0.0000 | 0.0000 | 0.0000 | 0.0001 | 0.0004 | 0.0009 | 0.0020 | 0.0041 | 0.0121 | 0.0321 | 0.0674 | 0.5690 | 0.1802 | 0.1317 |
| CCC | 0.0000 | 0.0000 | 0.0000 | 0.0000 | 0.0000 | 0.0000 | 0.0000 | 0.0000 | 0.0001 | 0.0002 | 0.0004 | 0.0008 | 0.0024 | 0.0064 | 0.0136 | 0.0413 | 0.6915 | 0.2433 |
| D | 0.0000 | 0.0000 | 0.0000 | 0.0000 | 0.0000 | 0.0000 | 0.0000 | 0.0000 | 0.0000 | 0.0000 | 0.0000 | 0.0000 | 0.0000 | 0.0000 | 0.0000 | 0.0000 | 0.0000 | 1.0000 |

**Figure 9d.  10-year transition probability matrix as of 1/1/2005 with 20-year half-life**

| Rating | AAA | AA1 | AA2 | AA3 | A1 | A2 | A3 | BAA1 | BAA2 | BAA3 | BA1 | BA2 | BA3 | B1 | B2 | B3 | CCC | D |
|---|---|---|---|---|---|---|---|---|---|---|---|---|---|---|---|---|---|---|
| AAA | 0.3895 | 0.1500 | 0.1386 | 0.1219 | 0.0842 | 0.0510 | 0.0286 | 0.0142 | 0.0098 | 0.0052 | 0.0025 | 0.0013 | 0.0011 | 0.0006 | 0.0004 | 0.0003 | 0.0003 | 0.0005 |
| AA1 | 0.1460 | 0.1211 | 0.1586 | 0.1783 | 0.1457 | 0.1002 | 0.0612 | 0.0324 | 0.0238 | 0.0133 | 0.0064 | 0.0035 | 0.0031 | 0.0018 | 0.0012 | 0.0009 | 0.0010 | 0.0016 |
| AA2 | 0.0664 | 0.0781 | 0.1395 | 0.1895 | 0.1748 | 0.1313 | 0.0850 | 0.0469 | 0.0359 | 0.0206 | 0.0102 | 0.0057 | 0.0051 | 0.0030 | 0.0020 | 0.0015 | 0.0018 | 0.0029 |
| AA3 | 0.0289 | 0.0435 | 0.0940 | 0.1718 | 0.1886 | 0.1598 | 0.1117 | 0.0653 | 0.0524 | 0.0312 | 0.0159 | 0.0091 | 0.0082 | 0.0050 | 0.0033 | 0.0026 | 0.0032 | 0.0054 |
| A1 | 0.0129 | 0.0230 | 0.0561 | 0.1221 | 0.1758 | 0.1775 | 0.1379 | 0.0871 | 0.0743 | 0.0464 | 0.0246 | 0.0144 | 0.0134 | 0.0083 | 0.0057 | 0.0045 | 0.0057 | 0.0103 |
| A2 | 0.0058 | 0.0117 | 0.0312 | 0.0764 | 0.1311 | 0.1724 | 0.1551 | 0.1084 | 0.1002 | 0.0664 | 0.0367 | 0.0223 | 0.0214 | 0.0137 | 0.0095 | 0.0078 | 0.0101 | 0.0198 |
| A3 | 0.0030 | 0.0065 | 0.0183 | 0.0486 | 0.0926 | 0.1410 | 0.1515 | 0.1195 | 0.1211 | 0.0858 | 0.0498 | 0.0314 | 0.0313 | 0.0207 | 0.0147 | 0.0124 | 0.0165 | 0.0354 |
| BAA1 | 0.0016 | 0.0036 | 0.0107 | 0.0301 | 0.0620 | 0.1044 | 0.1267 | 0.1177 | 0.1338 | 0.1029 | 0.0634 | 0.0419 | 0.0434 | 0.0298 | 0.0218 | 0.0188 | 0.0260 | 0.0614 |
| BAA2 | 0.0008 | 0.0021 | 0.0063 | 0.0186 | 0.0407 | 0.0743 | 0.0989 | 0.1031 | 0.1334 | 0.1119 | 0.0734 | 0.0509 | 0.0551 | 0.0394 | 0.0296 | 0.0261 | 0.0375 | 0.0977 |
| BAA3 | 0.0005 | 0.0012 | 0.0037 | 0.0113 | 0.0260 | 0.0503 | 0.0715 | 0.0809 | 0.1142 | 0.1081 | 0.0772 | 0.0573 | 0.0657 | 0.0495 | 0.0385 | 0.0351 | 0.0527 | 0.1563 |
| BA1 | 0.0002 | 0.0007 | 0.0021 | 0.0067 | 0.0158 | 0.0321 | 0.0479 | 0.0574 | 0.0863 | 0.0890 | 0.0717 | 0.0580 | 0.0715 | 0.0577 | 0.0467 | 0.0444 | 0.0702 | 0.2417 |
| BA2 | 0.0001 | 0.0004 | 0.0012 | 0.0038 | 0.0093 | 0.0195 | 0.0303 | 0.0381 | 0.0601 | 0.0663 | 0.0583 | 0.0526 | 0.0701 | 0.0609 | 0.0516 | 0.0510 | 0.0851 | 0.3412 |
| BA3 | 0.0001 | 0.0002 | 0.0006 | 0.0021 | 0.0053 | 0.0115 | 0.0184 | 0.0241 | 0.0398 | 0.0464 | 0.0439 | 0.0429 | 0.0627 | 0.0588 | 0.0521 | 0.0539 | 0.0947 | 0.4425 |
| B1 | 0.0000 | 0.0001 | 0.0003 | 0.0010 | 0.0025 | 0.0056 | 0.0093 | 0.0127 | 0.0217 | 0.0268 | 0.0270 | 0.0285 | 0.0450 | 0.0484 | 0.0460 | 0.0509 | 0.0972 | 0.5771 |
| B2 | 0.0000 | 0.0000 | 0.0002 | 0.0005 | 0.0014 | 0.0031 | 0.0052 | 0.0073 | 0.0129 | 0.0164 | 0.0173 | 0.0190 | 0.0315 | 0.0364 | 0.0374 | 0.0441 | 0.0909 | 0.6764 |
| B3 | 0.0000 | 0.0000 | 0.0001 | 0.0002 | 0.0006 | 0.0014 | 0.0025 | 0.0036 | 0.0065 | 0.0085 | 0.0093 | 0.0107 | 0.0184 | 0.0228 | 0.0250 | 0.0330 | 0.0763 | 0.7812 |
| CCC | 0.0000 | 0.0000 | 0.0000 | 0.0001 | 0.0002 | 0.0004 | 0.0008 | 0.0011 | 0.0021 | 0.0029 | 0.0034 | 0.0041 | 0.0074 | 0.0100 | 0.0118 | 0.0175 | 0.0512 | 0.8870 |
| D | 0.0000 | 0.0000 | 0.0000 | 0.0000 | 0.0000 | 0.0000 | 0.0000 | 0.0000 | 0.0000 | 0.0000 | 0.0000 | 0.0000 | 0.0000 | 0.0000 | 0.0000 | 0.0000 | 0.0000 | 1.0000 |

**Figure 10a. Gengen matrix estimated as of 1/1/2003 with 3-year half-life**

| Rating | AAA | AA1 | AA2 | AA3 | A1 | A2 | A3 | BAA1 | BAA2 | BAA3 | BA1 | BA2 | BA3 | B1 | B2 | B3 | CCC | D |
|---|---|---|---|---|---|---|---|---|---|---|---|---|---|---|---|---|---|---|
| AAA | (0.1936) | 0.1936 | - | - | - | - | - | - | - | - | - | - | - | - | - | - | - | - |
| AA1 | 0.2458 | (0.7909) | 0.5450 | - | - | - | - | - | - | - | - | - | - | - | - | - | - | - |
| AA2 | - | 0.3342 | (0.8632) | 0.5290 | - | - | - | - | - | - | - | - | - | - | - | - | - | - |
| AA3 | - | - | 0.2894 | (0.8003) | 0.5109 | - | - | - | - | - | - | - | - | - | - | - | - | - |
| A1 | - | - | - | 0.4482 | (1.0223) | 0.5741 | - | - | - | - | - | - | - | - | - | - | - | - |
| A2 | - | - | - | - | 0.3265 | (0.9193) | 0.5928 | - | - | - | - | - | - | - | - | - | - | - |
| A3 | - | - | - | - | - | 0.4738 | (1.2049) | 0.7311 | - | - | - | - | - | - | - | - | - | - |
| BAA1 | - | - | - | - | - | - | 0.8172 | (1.9334) | 1.1162 | - | - | - | - | - | - | - | - | - |
| BAA2 | - | - | - | - | - | - | - | 0.6918 | (1.4725) | 0.7807 | - | - | - | - | - | - | - | - |
| BAA3 | - | - | - | - | - | - | - | - | 0.6789 | (1.6572) | 0.9784 | - | - | - | - | - | - | - |
| BA1 | - | - | - | - | - | - | - | - | - | 1.0835 | (2.3856) | 1.3021 | - | - | - | - | - | - |
| BA2 | - | - | - | - | - | - | - | - | - | - | 1.1761 | (2.9586) | 1.7825 | - | - | - | - | - |
| BA3 | - | - | - | - | - | - | - | - | - | - | - | 1.1580 | (2.5782) | 1.4202 | - | - | - | - |
| B1 | - | - | - | - | - | - | - | - | - | - | - | - | 0.8427 | (2.4829) | 1.6402 | - | - | - |
| B2 | - | - | - | - | - | - | - | - | - | - | - | - | - | 0.9850 | (2.4838) | 1.4988 | - | - |
| B3 | - | - | - | - | - | - | - | - | - | - | - | - | - | - | 0.8522 | (2.5006) | 1.6484 | - |
| CCC | - | - | - | - | - | - | - | - | - | - | - | - | - | - | - | 0.2759 | (0.7694) | 0.4935 |
| D | - | - | - | - | - | - | - | - | - | - | - | - | - | - | - | - | - | - |

**Figure 10b. Transition generator matrix as of 1/1/2003 with 3-year half-life**

| Rating | AAA | AA1 | AA2 | AA3 | A1 | A2 | A3 | BAA1 | BAA2 | BAA3 | BA1 | BA2 | BA3 | B1 | B2 | B3 | CCC | D |
|---|---|---|---|---|---|---|---|---|---|---|---|---|---|---|---|---|---|---|
| AAA | (0.1419) | 0.0984 | 0.0303 | 0.0095 | 0.0025 | 0.0008 | 0.0002 | 0.0001 | 0.0000 | 0.0000 | 0.0000 | 0.0000 | 0.0000 | 0.0000 | 0.0000 | 0.0000 | 0.0000 | 0.0000 |
| AA1 | 0.1250 | (0.3933) | 0.1866 | 0.0587 | 0.0157 | 0.0051 | 0.0015 | 0.0004 | 0.0002 | 0.0001 | 0.0000 | 0.0000 | 0.0000 | 0.0000 | 0.0000 | 0.0000 | 0.0000 | 0.0000 |
| AA2 | 0.0236 | 0.1144 | (0.4005) | 0.1887 | 0.0503 | 0.0162 | 0.0049 | 0.0014 | 0.0007 | 0.0002 | 0.0001 | 0.0000 | 0.0000 | 0.0000 | 0.0000 | 0.0000 | 0.0000 | 0.0000 |
| AA3 | 0.0041 | 0.0197 | 0.1032 | (0.3726) | 0.1672 | 0.0540 | 0.0162 | 0.0046 | 0.0023 | 0.0008 | 0.0003 | 0.0001 | 0.0001 | 0.0000 | 0.0000 | 0.0000 | 0.0000 | 0.0000 |
| A1 | 0.0009 | 0.0046 | 0.0241 | 0.1467 | (0.4392) | 0.1812 | 0.0544 | 0.0154 | 0.0077 | 0.0026 | 0.0009 | 0.0003 | 0.0002 | 0.0001 | 0.0000 | 0.0000 | 0.0000 | 0.0000 |
| A2 | 0.0002 | 0.0008 | 0.0044 | 0.0270 | 0.1030 | (0.4040) | 0.1790 | 0.0505 | 0.0252 | 0.0086 | 0.0029 | 0.0011 | 0.0006 | 0.0003 | 0.0002 | 0.0001 | 0.0001 | 0.0000 |
| A3 | 0.0000 | 0.0002 | 0.0011 | 0.0065 | 0.0247 | 0.1431 | (0.4505) | 0.1552 | 0.0773 | 0.0263 | 0.0088 | 0.0035 | 0.0019 | 0.0009 | 0.0005 | 0.0002 | 0.0002 | 0.0001 |
| BAA1 | 0.0000 | 0.0001 | 0.0003 | 0.0020 | 0.0078 | 0.0452 | 0.1735 | (0.5648) | 0.2168 | 0.0737 | 0.0247 | 0.0097 | 0.0055 | 0.0026 | 0.0014 | 0.0006 | 0.0006 | 0.0003 |
| BAA2 | 0.0000 | 0.0000 | 0.0001 | 0.0006 | 0.0024 | 0.0139 | 0.0535 | 0.1343 | (0.4870) | 0.1745 | 0.0584 | 0.0230 | 0.0129 | 0.0062 | 0.0033 | 0.0015 | 0.0014 | 0.0007 |
| BAA3 | 0.0000 | 0.0000 | 0.0000 | 0.0002 | 0.0007 | 0.0041 | 0.0158 | 0.0397 | 0.1518 | (0.5125) | 0.1632 | 0.0642 | 0.0361 | 0.0173 | 0.0092 | 0.0042 | 0.0040 | 0.0020 |
| BA1 | 0.0000 | 0.0000 | 0.0000 | 0.0001 | 0.0003 | 0.0015 | 0.0059 | 0.0147 | 0.0563 | 0.1807 | (0.5974) | 0.1585 | 0.0890 | 0.0427 | 0.0227 | 0.0105 | 0.0098 | 0.0048 |
| BA2 | 0.0000 | 0.0000 | 0.0000 | 0.0000 | 0.0001 | 0.0005 | 0.0021 | 0.0052 | 0.0200 | 0.0642 | 0.1431 | (0.6414) | 0.2014 | 0.0966 | 0.0513 | 0.0237 | 0.0221 | 0.0109 |
| BA3 | 0.0000 | 0.0000 | 0.0000 | 0.0000 | 0.0000 | 0.0002 | 0.0008 | 0.0019 | 0.0073 | 0.0234 | 0.0522 | 0.1308 | (0.6115) | 0.1864 | 0.0990 | 0.0457 | 0.0426 | 0.0210 |
| B1 | 0.0000 | 0.0000 | 0.0000 | 0.0000 | 0.0000 | 0.0001 | 0.0002 | 0.0005 | 0.0021 | 0.0065 | 0.0149 | 0.0372 | 0.1106 | (0.6091) | 0.2075 | 0.0959 | 0.0893 | 0.0441 |
| B2 | 0.0000 | 0.0000 | 0.0000 | 0.0000 | 0.0000 | 0.0000 | 0.0001 | 0.0002 | 0.0007 | 0.0021 | 0.0047 | 0.0119 | 0.0353 | 0.1246 | (0.6102) | 0.1801 | 0.1678 | 0.0828 |
| B3 | 0.0000 | 0.0000 | 0.0000 | 0.0000 | 0.0000 | 0.0000 | 0.0000 | 0.0000 | 0.0002 | 0.0006 | 0.0012 | 0.0031 | 0.0093 | 0.0327 | 0.1024 | (0.6444) | 0.3313 | 0.1635 |
| CCC | 0.0000 | 0.0000 | 0.0000 | 0.0000 | 0.0000 | 0.0000 | 0.0000 | 0.0000 | 0.0000 | 0.0001 | 0.0002 | 0.0005 | 0.0014 | 0.0051 | 0.0160 | 0.0554 | (0.3832) | 0.3044 |
| D | - | - | - | - | - | - | - | - | - | - | - | - | - | - | - | - | - | - |

**Figure 10c. 1-year transition probability matrix as of 1/1/2003 with 3-year half-life**

| Rating | AAA | AA1 | AA2 | AA3 | A1 | A2 | A3 | BAA1 | BAA2 | BAA3 | BA1 | BA2 | BA3 | B1 | B2 | B3 | CCC | D |
|---|---|---|---|---|---|---|---|---|---|---|---|---|---|---|---|---|---|---|
| AAA | 0.8730 | 0.0773 | 0.0306 | 0.0124 | 0.0041 | 0.0016 | 0.0006 | 0.0002 | 0.0001 | 0.0000 | 0.0000 | 0.0000 | 0.0000 | 0.0000 | 0.0000 | 0.0000 | 0.0000 | 0.0000 |
| AA1 | 0.0982 | 0.6873 | 0.1304 | 0.0542 | 0.0181 | 0.0074 | 0.0027 | 0.0009 | 0.0005 | 0.0002 | 0.0001 | 0.0000 | 0.0000 | 0.0000 | 0.0000 | 0.0000 | 0.0000 | 0.0000 |
| AA2 | 0.0238 | 0.0799 | 0.6850 | 0.1349 | 0.0457 | 0.0189 | 0.0070 | 0.0023 | 0.0014 | 0.0006 | 0.0002 | 0.0001 | 0.0001 | 0.0000 | 0.0000 | 0.0000 | 0.0000 | 0.0000 |
| AA3 | 0.0053 | 0.0182 | 0.0738 | 0.7055 | 0.1168 | 0.0492 | 0.0186 | 0.0062 | 0.0037 | 0.0015 | 0.0006 | 0.0003 | 0.0002 | 0.0001 | 0.0001 | 0.0000 | 0.0000 | 0.0000 |
| A1 | 0.0015 | 0.0053 | 0.0219 | 0.1024 | 0.6603 | 0.1263 | 0.0484 | 0.0165 | 0.0099 | 0.0041 | 0.0016 | 0.0007 | 0.0004 | 0.0002 | 0.0001 | 0.0001 | 0.0001 | 0.0001 |
| A2 | 0.0003 | 0.0012 | 0.0052 | 0.0246 | 0.0718 | 0.6845 | 0.1239 | 0.0427 | 0.0261 | 0.0108 | 0.0042 | 0.0019 | 0.0012 | 0.0007 | 0.0004 | 0.0002 | 0.0002 | 0.0002 |
| A3 | 0.0001 | 0.0004 | 0.0015 | 0.0074 | 0.0220 | 0.0990 | 0.6570 | 0.1012 | 0.0629 | 0.0264 | 0.0103 | 0.0047 | 0.0030 | 0.0017 | 0.0010 | 0.0005 | 0.0006 | 0.0004 |
| BAA1 | 0.0000 | 0.0001 | 0.0006 | 0.0028 | 0.0084 | 0.0381 | 0.1131 | 0.5880 | 0.1391 | 0.0591 | 0.0234 | 0.0107 | 0.0069 | 0.0039 | 0.0023 | 0.0012 | 0.0014 | 0.0010 |
| BAA2 | 0.0000 | 0.0000 | 0.0002 | 0.0010 | 0.0031 | 0.0144 | 0.0435 | 0.0862 | 0.6353 | 0.1153 | 0.0461 | 0.0212 | 0.0138 | 0.0077 | 0.0047 | 0.0025 | 0.0028 | 0.0020 |
| BAA3 | 0.0000 | 0.0000 | 0.0001 | 0.0004 | 0.0011 | 0.0052 | 0.0159 | 0.0319 | 0.1003 | 0.6196 | 0.1019 | 0.0474 | 0.0311 | 0.0176 | 0.0108 | 0.0057 | 0.0065 | 0.0046 |
| BA1 | 0.0000 | 0.0000 | 0.0000 | 0.0002 | 0.0005 | 0.0022 | 0.0069 | 0.0140 | 0.0444 | 0.1128 | 0.5688 | 0.0946 | 0.0627 | 0.0358 | 0.0221 | 0.0117 | 0.0136 | 0.0098 |
| BA2 | 0.0000 | 0.0000 | 0.0000 | 0.0001 | 0.0002 | 0.0009 | 0.0028 | 0.0058 | 0.0185 | 0.0474 | 0.0855 | 0.5434 | 0.1175 | 0.0678 | 0.0422 | 0.0225 | 0.0263 | 0.0192 |
| BA3 | 0.0000 | 0.0000 | 0.0000 | 0.0000 | 0.0001 | 0.0004 | 0.0012 | 0.0024 | 0.0078 | 0.0202 | 0.0368 | 0.0764 | 0.5592 | 0.1114 | 0.0698 | 0.0376 | 0.0442 | 0.0327 |
| B1 | 0.0000 | 0.0000 | 0.0000 | 0.0000 | 0.0000 | 0.0001 | 0.0004 | 0.0008 | 0.0026 | 0.0068 | 0.0125 | 0.0261 | 0.0661 | 0.5598 | 0.1215 | 0.0659 | 0.0784 | 0.0590 |
| B2 | 0.0000 | 0.0000 | 0.0000 | 0.0000 | 0.0000 | 0.0000 | 0.0001 | 0.0003 | 0.0009 | 0.0025 | 0.0046 | 0.0098 | 0.0249 | 0.0729 | 0.5581 | 0.1042 | 0.1255 | 0.0962 |
| B3 | 0.0000 | 0.0000 | 0.0000 | 0.0000 | 0.0000 | 0.0000 | 0.0000 | 0.0001 | 0.0003 | 0.0007 | 0.0014 | 0.0030 | 0.0076 | 0.0225 | 0.0592 | 0.5367 | 0.2065 | 0.1619 |
| CCC | 0.0000 | 0.0000 | 0.0000 | 0.0000 | 0.0000 | 0.0000 | 0.0000 | 0.0000 | 0.0001 | 0.0001 | 0.0003 | 0.0006 | 0.0015 | 0.0045 | 0.0119 | 0.0346 | 0.6888 | 0.2577 |
| D | 0.0000 | 0.0000 | 0.0000 | 0.0000 | 0.0000 | 0.0000 | 0.0000 | 0.0000 | 0.0000 | 0.0000 | 0.0000 | 0.0000 | 0.0000 | 0.0000 | 0.0000 | 0.0000 | 0.0000 | 1.0000 |

**Figure 10d. 10-year transition probability matrix as of 1/1/2003 with 3-year half-life**

| Rating | AAA | AA1 | AA2 | AA3 | A1 | A2 | A3 | BAA1 | BAA2 | BAA3 | BA1 | BA2 | BA3 | B1 | B2 | B3 | CCC | D |
|---|---|---|---|---|---|---|---|---|---|---|---|---|---|---|---|---|---|---|
| AAA | 0.3598 | 0.1525 | 0.1363 | 0.1235 | 0.0740 | 0.0575 | 0.0361 | 0.0179 | 0.0160 | 0.0096 | 0.0049 | 0.0028 | 0.0023 | 0.0016 | 0.0012 | 0.0008 | 0.0013 | 0.0020 |
| AA1 | 0.1937 | 0.1246 | 0.1424 | 0.1569 | 0.1059 | 0.0924 | 0.0629 | 0.0329 | 0.0310 | 0.0194 | 0.0102 | 0.0061 | 0.0050 | 0.0037 | 0.0028 | 0.0019 | 0.0030 | 0.0052 |
| AA2 | 0.1061 | 0.0873 | 0.1276 | 0.1669 | 0.1241 | 0.1186 | 0.0857 | 0.0467 | 0.0456 | 0.0295 | 0.0159 | 0.0096 | 0.0081 | 0.0060 | 0.0047 | 0.0031 | 0.0052 | 0.0094 |
| AA3 | 0.0526 | 0.0526 | 0.0913 | 0.1534 | 0.1301 | 0.1405 | 0.1099 | 0.0629 | 0.0643 | 0.0432 | 0.0238 | 0.0148 | 0.0126 | 0.0096 | 0.0076 | 0.0051 | 0.0087 | 0.0169 |
| A1 | 0.0277 | 0.0312 | 0.0595 | 0.1141 | 0.1164 | 0.1482 | 0.1281 | 0.0783 | 0.0845 | 0.0596 | 0.0338 | 0.0215 | 0.0188 | 0.0147 | 0.0118 | 0.0081 | 0.0142 | 0.0295 |
| A2 | 0.0122 | 0.0155 | 0.0324 | 0.0701 | 0.0843 | 0.1390 | 0.1379 | 0.0917 | 0.1061 | 0.0792 | 0.0466 | 0.0306 | 0.0275 | 0.0220 | 0.0181 | 0.0127 | 0.0228 | 0.0516 |
| A3 | 0.0061 | 0.0084 | 0.0187 | 0.0438 | 0.0582 | 0.1102 | 0.1284 | 0.0940 | 0.1177 | 0.0936 | 0.0574 | 0.0390 | 0.0360 | 0.0297 | 0.0250 | 0.0179 | 0.0332 | 0.0826 |
| BAA1 | 0.0034 | 0.0049 | 0.0114 | 0.0280 | 0.0398 | 0.0819 | 0.1051 | 0.0867 | 0.1186 | 0.1015 | 0.0651 | 0.0459 | 0.0438 | 0.0374 | 0.0322 | 0.0237 | 0.0455 | 0.1252 |
| BAA2 | 0.0019 | 0.0029 | 0.0069 | 0.0178 | 0.0266 | 0.0588 | 0.0815 | 0.0735 | 0.1112 | 0.1027 | 0.0690 | 0.0507 | 0.0499 | 0.0440 | 0.0389 | 0.0293 | 0.0581 | 0.1765 |
| BAA3 | 0.0010 | 0.0016 | 0.0039 | 0.0104 | 0.0163 | 0.0381 | 0.0564 | 0.0547 | 0.0893 | 0.0927 | 0.0667 | 0.0519 | 0.0537 | 0.0497 | 0.0456 | 0.0356 | 0.0739 | 0.2586 |
| BA1 | 0.0006 | 0.0009 | 0.0023 | 0.0063 | 0.0103 | 0.0249 | 0.0383 | 0.0388 | 0.0664 | 0.0738 | 0.0576 | 0.0480 | 0.0523 | 0.0513 | 0.0489 | 0.0397 | 0.0867 | 0.3529 |
| BA2 | 0.0003 | 0.0005 | 0.0013 | 0.0036 | 0.0059 | 0.0147 | 0.0235 | 0.0248 | 0.0441 | 0.0520 | 0.0434 | 0.0398 | 0.0465 | 0.0488 | 0.0489 | 0.0415 | 0.0963 | 0.4642 |
| BA3 | 0.0002 | 0.0003 | 0.0007 | 0.0020 | 0.0034 | 0.0086 | 0.0141 | 0.0153 | 0.0282 | 0.0349 | 0.0307 | 0.0302 | 0.0383 | 0.0433 | 0.0456 | 0.0406 | 0.0995 | 0.5643 |
| B1 | 0.0001 | 0.0001 | 0.0003 | 0.0009 | 0.0016 | 0.0041 | 0.0069 | 0.0078 | 0.0148 | 0.0192 | 0.0178 | 0.0188 | 0.0257 | 0.0330 | 0.0375 | 0.0358 | 0.0953 | 0.6805 |
| B2 | 0.0000 | 0.0001 | 0.0001 | 0.0004 | 0.0007 | 0.0020 | 0.0035 | 0.0040 | 0.0078 | 0.0106 | 0.0102 | 0.0113 | 0.0162 | 0.0225 | 0.0282 | 0.0290 | 0.0844 | 0.7689 |
| B3 | 0.0000 | 0.0000 | 0.0001 | 0.0002 | 0.0003 | 0.0008 | 0.0014 | 0.0017 | 0.0034 | 0.0047 | 0.0047 | 0.0055 | 0.0082 | 0.0122 | 0.0165 | 0.0199 | 0.0658 | 0.8547 |
| CCC | 0.0000 | 0.0000 | 0.0000 | 0.0000 | 0.0001 | 0.0002 | 0.0004 | 0.0005 | 0.0011 | 0.0016 | 0.0017 | 0.0021 | 0.0034 | 0.0054 | 0.0080 | 0.0110 | 0.0454 | 0.9188 |
| D | 0.0000 | 0.0000 | 0.0000 | 0.0000 | 0.0000 | 0.0000 | 0.0000 | 0.0000 | 0.0000 | 0.0000 | 0.0000 | 0.0000 | 0.0000 | 0.0000 | 0.0000 | 0.0000 | 0.0000 | 1.0000 |

**Figure 11a. Moody's idealized cumulative default probabilities and WARFs**

| Horizon (yrs) | 1 | 2 | 3 | 4 | 5 | 6 | 7 | 8 | 9 | 10 | WARF |
|---|---|---|---|---|---|---|---|---|---|---|---|
| Aaa | 0.0001% | 0.0002% | 0.0007% | 0.0018% | 0.0029% | 0.0040% | 0.0052% | 0.0066% | 0.0082% | 0.0100% | 1 |
| Aa1 | 0.0006% | 0.0030% | 0.0100% | 0.0210% | 0.0310% | 0.0420% | 0.0540% | 0.0670% | 0.0820% | 0.1000% | 10 |
| Aa2 | 0.0014% | 0.0080% | 0.0260% | 0.0470% | 0.0680% | 0.0890% | 0.1110% | 0.1350% | 0.1640% | 0.2000% | 20 |
| Aa3 | 0.0030% | 0.0190% | 0.0590% | 0.1010% | 0.1420% | 0.1830% | 0.2270% | 0.2720% | 0.3270% | 0.4000% | 40 |
| A1 | 0.0058% | 0.0370% | 0.1170% | 0.1890% | 0.2610% | 0.3300% | 0.4060% | 0.4800% | 0.5730% | 0.7000% | 70 |
| A2 | 0.0109% | 0.0700% | 0.2220% | 0.3450% | 0.4670% | 0.5830% | 0.7100% | 0.8290% | 0.9820% | 1.2000% | 120 |
| A3 | 0.0389% | 0.1500% | 0.3600% | 0.5400% | 0.7300% | 0.9100% | 1.1100% | 1.3000% | 1.5200% | 1.8000% | 180 |
| Baa1 | 0.0900% | 0.2800% | 0.5600% | 0.8300% | 1.1000% | 1.3700% | 1.6700% | 1.9700% | 2.2700% | 2.6000% | 260 |
| Baa2 | 0.1700% | 0.4700% | 0.8300% | 1.2000% | 1.5800% | 1.9700% | 2.4100% | 2.8500% | 3.2400% | 3.6000% | 360 |
| Baa3 | 0.4200% | 1.0500% | 1.7100% | 2.3800% | 3.0500% | 3.7000% | 4.3300% | 4.9700% | 5.5700% | 6.1000% | 610 |
| Ba1 | 0.8700% | 2.0200% | 3.1300% | 4.2000% | 5.2800% | 6.2500% | 7.0600% | 7.8900% | 8.6900% | 9.4000% | 940 |
| Ba2 | 1.5600% | 3.4700% | 5.1800% | 6.8000% | 8.4100% | 9.7700% | 10.7000% | 11.6600% | 12.6500% | 13.5000% | 1350 |
| Ba3 | 2.8100% | 5.5100% | 7.8700% | 9.7900% | 11.8600% | 13.4900% | 14.6200% | 15.7100% | 16.7100% | 17.6600% | 1766 |
| B1 | 4.6800% | 8.3800% | 11.5800% | 13.8500% | 16.1200% | 17.8900% | 19.1300% | 20.2300% | 21.2400% | 22.2000% | 2220 |
| B2 | 7.1600% | 11.6700% | 15.5500% | 18.1300% | 20.7100% | 22.6500% | 24.0100% | 25.1500% | 26.2200% | 27.2000% | 2720 |
| B3 | 11.6200% | 16.6100% | 21.0300% | 24.0400% | 27.0500% | 29.2000% | 31.0000% | 32.5800% | 33.7800% | 34.9000% | 3490 |
| Caa | 26.0000% | 32.5000% | 39.0000% | 43.8800% | 48.7500% | 52.0000% | 55.2500% | 58.5000% | 61.7500% | 65.0000% | 6500 |

**Figure 11b. Model-based cumulative default probabilities, estimated as of 1/1/2005 with 20-year half-life**

| Horizon (yrs) | 1 | 2 | 3 | 4 | 5 | 6 | 7 | 8 | 9 | 10 | WARF |
|---|---|---|---|---|---|---|---|---|---|---|---|
| Aaa | 0.0000% | 0.0001% | 0.0003% | 0.0009% | 0.0022% | 0.0049% | 0.0096% | 0.0176% | 0.0299% | 0.0483% | 1 |
| Aa1 | 0.0000% | 0.0004% | 0.0014% | 0.0041% | 0.0096% | 0.0198% | 0.0370% | 0.0637% | 0.1031% | 0.1583% | 10 |
| Aa2 | 0.0001% | 0.0009% | 0.0034% | 0.0093% | 0.0211% | 0.0417% | 0.0749% | 0.1246% | 0.1952% | 0.2911% | 20 |
| Aa3 | 0.0004% | 0.0025% | 0.0086% | 0.0221% | 0.0474% | 0.0896% | 0.1542% | 0.2472% | 0.3741% | 0.5403% | 40 |
| A1 | 0.0012% | 0.0071% | 0.0227% | 0.0548% | 0.1112% | 0.2001% | 0.3295% | 0.5070% | 0.7391% | 1.0312% | 70 |
| A2 | 0.0040% | 0.0206% | 0.0610% | 0.1380% | 0.2645% | 0.4523% | 0.7115% | 1.0500% | 1.4731% | 1.9836% | 120 |
| A3 | 0.0117% | 0.0545% | 0.1497% | 0.3178% | 0.5765% | 0.9391% | 1.4143% | 2.0061% | 2.7146% | 3.5359% | 180 |
| Baa1 | 0.0329% | 0.1394% | 0.3551% | 0.7083% | 1.2174% | 1.8911% | 2.7295% | 3.7256% | 4.8674% | 6.1392% | 260 |
| Baa2 | 0.0809% | 0.3130% | 0.7448% | 1.4046% | 2.3010% | 3.4271% | 4.7647% | 6.2881% | 7.9682% | 9.7743% | 360 |
| Baa3 | 0.2108% | 0.7360% | 1.6211% | 2.8691% | 4.4524% | 6.3252% | 8.4335% | 10.7210% | 13.1350% | 15.6280% | 610 |
| Ba1 | 0.5259% | 1.6594% | 3.3864% | 5.6303% | 8.2854% | 11.2400% | 14.3880% | 17.6400% | 20.9200% | 24.1700% | 940 |
| Ba2 | 1.1218% | 3.2352% | 6.1724% | 9.7186% | 13.6620% | 17.8180% | 22.0380% | 26.2110% | 30.2570% | 34.1230% | 1350 |
| Ba3 | 2.0953% | 5.5632% | 9.9847% | 14.9730% | 20.2150% | 25.4760% | 30.5950% | 35.4650% | 40.0260% | 44.2500% | 1766 |
| B1 | 4.3881% | 10.4460% | 17.2880% | 24.3100% | 31.1320% | 37.5370% | 43.4140% | 48.7280% | 53.4830% | 57.7100% | 2220 |
| B2 | 7.5174% | 16.2980% | 25.2020% | 33.6270% | 41.2960% | 48.1130% | 54.0880% | 59.2790% | 63.7650% | 67.6350% | 2720 |
| B3 | 13.1730% | 25.7310% | 36.8700% | 46.4040% | 54.4110% | 61.0740% | 66.5960% | 71.1700% | 74.9650% | 78.1210% | 3490 |
| Caa | 24.3320% | 41.8370% | 54.6300% | 64.1150% | 71.2430% | 76.6690% | 80.8470% | 84.1000% | 86.6610% | 88.6960% | 6500 |

**Figure 11c. Model-based cumulative default probabilities, estimated as of 1/1/2003 with 3-year half-life**

| Horizon (yrs) | 1 | 2 | 3 | 4 | 5 | 6 | 7 | 8 | 9 | 10 | WARF |
|---|---|---|---|---|---|---|---|---|---|---|---|
| Aaa | 0.0001% | 0.0004% | 0.0016% | 0.0048% | 0.0115% | 0.0240% | 0.0453% | 0.0790% | 0.1291% | 0.2002% | 1 |
| Aa1 | 0.0003% | 0.0018% | 0.0065% | 0.0175% | 0.0393% | 0.0771% | 0.1373% | 0.2266% | 0.3520% | 0.5206% | 10 |
| Aa2 | 0.0007% | 0.0044% | 0.0153% | 0.0394% | 0.0843% | 0.1588% | 0.2722% | 0.4342% | 0.6538% | 0.9391% | 20 |
| Aa3 | 0.0020% | 0.0114% | 0.0369% | 0.0895% | 0.1821% | 0.3282% | 0.5407% | 0.8316% | 1.2107% | 1.6860% | 40 |
| A1 | 0.0055% | 0.0292% | 0.0878% | 0.2007% | 0.3874% | 0.6661% | 1.0521% | 1.5571% | 2.1889% | 2.9510% | 70 |
| A2 | 0.0154% | 0.0742% | 0.2077% | 0.4470% | 0.8194% | 1.3457% | 2.0396% | 2.9072% | 3.9479% | 5.1554% | 120 |
| A3 | 0.0399% | 0.1737% | 0.4515% | 0.9138% | 1.5889% | 2.4910% | 3.6220% | 4.9731% | 6.5278% | 8.2640% | 180 |
| Baa1 | 0.0951% | 0.3770% | 0.9125% | 1.7420% | 2.8804% | 4.3213% | 6.0416% | 8.0082% | 10.1820% | 12.5210% | 260 |
| Baa2 | 0.1962% | 0.7183% | 1.6369% | 2.9745% | 4.7154% | 6.8178% | 9.2250% | 11.8740% | 14.7010% | 17.6470% | 360 |
| Baa3 | 0.4630% | 1.5334% | 3.2390% | 5.5307% | 8.3153% | 11.4810% | 14.9150% | 18.5130% | 22.1860% | 25.8620% | 610 |
| Ba1 | 0.9771% | 2.9454% | 5.7970% | 9.3459% | 13.3860% | 17.7260% | 22.2000% | 26.6780% | 31.0650% | 35.2910% | 940 |
| Ba2 | 1.9191% | 5.3031% | 9.7765% | 14.9480% | 20.4760% | 26.0920% | 31.6020% | 36.8740% | 41.8280% | 46.4220% | 1350 |
| Ba3 | 3.2681% | 8.3637% | 14.5680% | 21.2890% | 28.0930% | 34.6850% | 40.8840% | 46.5950% | 51.7770% | 56.4320% | 1766 |
| B1 | 5.9028% | 13.7440% | 22.3250% | 30.8780% | 38.9540% | 46.3240% | 52.9000% | 58.6790% | 63.7070% | 68.0510% | 2220 |
| B2 | 9.6154% | 20.4370% | 31.0720% | 40.8420% | 49.4790% | 56.9380% | 63.2870% | 68.6400% | 73.1310% | 76.8870% | 2720 |
| B3 | 16.1860% | 30.9300% | 43.5250% | 53.9410% | 62.4080% | 69.2320% | 74.7100% | 79.1040% | 82.6320% | 85.4720% | 3490 |
| Caa | 25.7670% | 44.2220% | 57.6170% | 67.4570% | 74.7650% | 80.2440% | 84.3920% | 87.5570% | 89.9930% | 91.8810% | 6500 |

**Figure 12.  Model-based cumulative probabilities of downgrade to high yield, estimated as of 1/1/2003 with 3-year half-life**

| Horizon (yrs) | 1 | 2 | 3 | 4 | 5 | 6 | 7 | 8 | 9 | 10 |
|---|---|---|---|---|---|---|---|---|---|---|
| Aaa | 0.0010% | 0.0057% | 0.0180% | 0.0434% | 0.0883% | 0.1595% | 0.2645% | 0.4103% | 0.6038% | 0.8509% |
| Aa1 | 0.0062% | 0.0300% | 0.0848% | 0.1852% | 0.3452% | 0.5774% | 0.8919% | 1.2966% | 1.7963% | 2.3936% |
| Aa2 | 0.0172% | 0.0744% | 0.1944% | 0.3976% | 0.7010% | 1.1167% | 1.6521% | 2.3103% | 3.0906% | 3.9889% |
| Aa3 | 0.0493% | 0.1903% | 0.4554% | 0.8673% | 1.4385% | 2.1726% | 3.0661% | 4.1104% | 5.2930% | 6.5994% |
| A1 | 0.1504% | 0.5107% | 1.1101% | 1.9538% | 3.0299% | 4.3164% | 5.7854% | 7.4072% | 9.1524% | 10.9940% |
| A2 | 0.4672% | 1.3861% | 2.7201% | 4.4045% | 6.3653% | 8.5312% | 10.8390% | 13.2350% | 15.6770% | 18.1300% |
| A3 | 1.3160% | 3.4052% | 6.0313% | 8.9928% | 12.1330% | 15.3380% | 18.5280% | 21.6510% | 24.6720% | 27.5750% |
| Baa1 | 3.5620% | 8.0123% | 12.7800% | 17.5300% | 22.0850% | 26.3560% | 30.3140% | 33.9590% | 37.3070% | 40.3810% |
| Baa2 | 8.3915% | 16.5170% | 23.8710% | 30.3300% | 35.9330% | 40.7790% | 44.9810% | 48.6420% | 51.8530% | 54.6900% |
| Baa3 | 20.7460% | 34.7790% | 44.6370% | 51.8240% | 57.2510% | 61.4820% | 64.8740% | 67.6610% | 69.9990% | 71.9940% |

Finally, in Figure 12 we show the cumulative downgrade probabilities for various horizons based on the short-term estimate as of 1/1/2003. To obtain the downgrade estimate, we first must modify the transition generator matrix, trimming it down to investment-grade-only ratings, and adding an additional row and column that stands for "High Yield or default" state. We then designate this state an absorbing state by setting its corresponding row of the downgrade generator matrix to zero.

This corresponds to counting a "first passage time" of the downgrade to high yield. Indeed, although a company can of course be downgraded to high yield and then subsequently upgraded back to investment grade, there can be many situations when the downgrade would trigger certain contingent events, such as a call of the company's bonds, or drawdown on a revolver loan facility, or a substantial increase in margin requirements. The cumulative downgrade probabilities calculated in a manner described above will be particularly useful in valuation and risk management of such ratings-contingent claims.

### Forecasting Transition Probabilities

Ultimately, investors are rarely interested in description of the past. Even when the question is "what was the downgrade probability in the past three years" it is often implicit that the investor will proceed to use such an estimate for some sort of forecast. Strictly speaking, there is no basis for such use, unless one can indeed prove that the historically estimated transition probabilities have explanatory power for future observations. We will discuss these questions briefly in the latter part of this section, after presenting the results of rolling estimates of historical transition probabilities.

In an earlier paper (Berd and Voronov [2002]) we investigated the continuous-time estimation of credit rating transition matrices and considered multi-factor forecasting models using both macro and industry variables. Our results, summarized in Figure 13 (which were obtained in a setting with a sharp window cut-off and without the parametric smoothing introduced in this paper) indicate that macro-economic variables such as capacity utilization (which reflects the business cycle) and swap spread (which is meant to reflect the liquidity crises), do carry a substantial explanatory power for observed rating transitions. Even more explanatory power is associated with industry factors – the improvement of the fit due to distinguishing between broad industry sectors such as telecoms-media-technology, basic industries, banks-and-brokerages and utilities is substantially greater than the improvement achieved by one or even two macro factors. Finally, combining the industry indicative factors with the business cycle macro factor produces the best fit among the considered models.

**Figure 13.   Comparative performance of explanatory variables**

| Model Specification | | | |
|---|---|---|---|
| Constant | Macro Factors | Industry Factors | Log-Likelihood |
| C | - | | -22016 |
| C | 1-factor: Capacity Utilization | | -21836 |
| C | 2-factor: Capacity Utilization, 5-yr Swap Spread | | -21626 |
| C | - | 9 Indicative Factors | -21475 |
| C | 1-factor: Capacity Utilization | 9 Indicative Factors | -21305 |

*Source: Lehman Brothers*

We expect that these stylized facts about the predictability of rating transitions will remain in effect after we update these results using the time-weighted parametrically smoothed estimation technique introduced here. The results of that investigation will be reported in a forthcoming paper.

The importance of time variation of hazard rates as functions of a firm's distance to default (which in our case is approximated by the rating process itself) and macro-economic variables such as U.S. income growth (which is similar in properties to our choice of capacity utilization) was recently investigated by Duffie and Wang (2004). They conclude that the macro-economic variables are important even after the issuer-specific information (distance to default) has been taken into account. Interestingly enough, while the evidence of such time-varying systemic credit risk is strong, the resulting pattern of defaults does not provide compelling evidence for default clustering (i.e. correlation between defaults) beyond that which can be ascribed to higher hazard rates with independent arrival times, as shown by Das, Duffie and Kapadia (2004).

## CONCLUSIONS

In this paper we have developed a consistent and robust methodology for estimation of credit rating transition probabilities. Building on the earlier work by Lando et al. and others, we have further extended the continuous-time Markov rating transition formalism by adopting a time-weighted estimation technique for the log-likelihood function and the parametric smoothing of transition generator matrix using the "hidden rating process" approximation. This allowed us to reduce the number of model parameters and achieve a higher accuracy in empirical estimation even for 3- and 5-year half-life estimates. Appendix B presents a simple technique for coarse-graining our notched-rating transition probability estimates in order to produce letter-grade transition matrices.

Some of the uses of transition matrix models include:

- Credit portfolio loss estimates for diversified portfolios pooled by rating. As a particular example, many CLOs report the rating composition of the collateral portfolios but not the specific issuers, in which case our model can be used for estimating the (real-world, not implied) loss rates for the collateral and the tranches.

- Valuation and risk management of rating-contingent obligations, such as bonds with step-up coupon provisions.

- Risk management of revolver loan commitments. It is often assumed that a credit line will be drawn if an issuer is downgraded to high yield and is no longer able to access the capital markets at advantageous rates. Our cumulative downgrade probability estimates can be used for scenario-based risk management of revolvers.

- Counterparty risk management. Many smaller trading counterparties do not have any outstanding public debt or CDS (or even equity) on which to base the estimates of their credit risk and margin requirements. Given credit analyst assessment of "shadow" rating of such counterparties, one can use the transition matrix model for setting and monitoring margin requirements and lines of credit.

We hope that our model will be useful for investors in these and other applications.

*Acknowledgments:* I would like to thank Artem Voronov (NYU) for research assistance, and Lea Carty, Roy Mashal and Marco Naldi for valuable comments and discussions.

## APPENDIX A: LETTER-GRADE RATINGS AND PARTIAL CENSORING

Prior to 1992, Moody's assigned only letter-grade credit ratings (Aaa, Aa, A, Baa, Ba, B, Caa) to issuers. In January 1992, the agency switched to a notched rating system, by splitting each of the letter grade ratings (except Aaa) into three alpha-numeric ratings (for example, the Baa category was split into Baa1, Baa2 and Baa3).

This change of rating denominations presents a challenge for empirical estimation. We confront it by regarding the pre-1992 letter-grade ratings as partially censored versions of the full "unobservable" notched rating system.

For any transition from a letter-grade rating R to a notched rating $r$, we set the mini-log-likelihood function to equally weighted average of the values corresponding to transitions from "unobserved" notched ratings R1, R2, and R3 to the final rating, which is consistent with having no information on the unobserved notch state prior to transition. Similarly, a transition from a notched rating to a letter-grade rating (if such is ever observed) is treated as a partially censored observation corresponding to either one of the "unobserved" final notched ratings. Finally, the pre-1992 transitions from letter-grade to letter-grade rating are treated as partially censored both in the beginning and at the end of the observation.

These rules are summarized in the following set of equations:

$$[22] \begin{cases} L_t\left(k, R_k^s, r_k^e\right) &= \frac{1}{3} \cdot \left[L_t\left(k, R1_k^s, r_k^e\right) + L_t\left(k, R2_k^s, r_k^e\right) + L_t\left(k, R3_k^s, r_k^e\right)\right] \\ L_t\left(k, r_k^s, R_k^e\right) &= \frac{1}{3} \cdot \left[L_t\left(k, r_k^s, R1_k^e\right) + L_t\left(k, r_k^s, R2_k^e\right) + L_t\left(k, r_k^s, R3_k^e\right)\right] \\ L_t\left(k, R_k^s, R_k^e\right) &= \frac{1}{9} \cdot \left[L_t\left(k, R1_k^s, R1_k^e\right) + L_t\left(k, R2_k^s, R1_k^e\right) + L_t\left(k, R3_k^s, R1_k^e\right) \right. \\ &+ \quad L_t\left(k, R1_k^s, R2_k^e\right) + L_t\left(k, R2_k^s, R2_k^e\right) + L_t\left(k, R3_k^s, R2_k^e\right) \\ &+ \quad \left. L_t\left(k, R1_k^s, R3_k^e\right) + L_t\left(k, R2_k^s, R3_k^e\right) + L_t\left(k, R3_k^s, R3_k^e\right)\right] \end{cases}$$

Note that in a situation when the observed transition is from a letter grade to one of its subset notched ratings (such as from Ba to Ba2) then one of the possible "unobserved" initial states is the same as the final state, and therefore corresponds to a "no-transition" event. This is automatically taken into account by properly calculating the mini-log-likelihood function for such events, which does not contain the second non-integral part [16]. A similar statement is valid for a transition from a notched rating to a letter-grade rating.

We have also made a simplification by grouping all Caa1 or below ratings into a single Caa designation. We do not make any adjustments for censored observations corresponding to transitions between these states.

Substituting these terms into the log-likelihood function [21] we obtain the necessary objective function that covers the entire period available in the Moody's database.

### APPENDIX B: COARSE-GRAINING THE TRANSITION MATRICES

Investors often ask for transition probability estimates on a coarse-grained, letter-grade scale spanning eight rating states [Aaa, Aa, A, Baa, Ba, B, Caa, D]. This can be useful when addressing macro trends in credit markets or broad asset allocations strategies, where it is impractical to maintain very fine classification with 18 notched rating states [Aaa, Aa1, Aa2, Aa3, A1, A2, A3, Baa1, Baa2, Baa3, Ba1, Ba2, Ba3, B1, B2, B3, Caa, D].

One could potentially answer this question by re-mapping all transition events to their corresponding letter-grade ratings and then applying the same estimation methodology as described in this paper with a smaller set of ratings. The problem with this approach is that it would lead to a dramatic reduction in the number of observed transitions and potentially less robust estimation of the model parameters.

The alternative is to start with the already estimated transition generator matrix for the notched ratings and derive a generator matrix for a coarse-grained rating variable that continues to satisfy the Markov property.

Let $R^s$ denote the coarse-grained initial rating state which has three notched sub-ratings $R1^s, R2^s, R3^s$. Let also $r^e$ denote a final state which has no sub-ratings, i.e. one of [Aaa, Caa, D]. Assuming equally probable initial sub-ratings, the transition intensity from $R^s$ to $r^e$ is equal to the average of transition intensities from initial sub-ratings to the final state.

Let now $R^e$ denote the coarse-grained final rating state and $r^s$ denote a initial state which has no sub-ratings. Since the final notched states are mutually exclusive, the transition intensity from $r^s$ to $R^e$ is equal to the sum of transition intensities from the initial state to the final sub-ratings.

Finally, the transition intensity between two coarse-grained states can be obtained by applying the above rules iteratively. These rules are summarized as follows:

$$[23] \quad \begin{cases} \lambda\left(R^s, r^e\right) &= \dfrac{1}{3} \cdot \left[\lambda\left(R1^s, r^e\right) + \lambda\left(R2^s, r^e\right) + \lambda\left(R3^s, r^e\right)\right] \\ \lambda\left(r^s, R^e\right) &= \lambda\left(r^s, R1^e\right) + \lambda\left(r^s, R2^e\right) + \lambda\left(r^s, R3^e\right) \\ \lambda\left(R^s, R^e\right) &= \dfrac{1}{3} \cdot \left[\lambda\left(R1^s, R1^e\right) + \lambda\left(R2^s, R1^e\right) + \lambda\left(R3^s, R1^e\right)\right. \\ &\quad + \lambda\left(R1^s, R2^e\right) + \lambda\left(R2^s, R2^e\right) + \lambda\left(R3^s, R2^e\right) \\ &\quad + \left.\lambda\left(R1^s, R3^e\right) + \lambda\left(R2^s, R3^e\right) + \lambda\left(R3^s, R3^e\right)\right] \end{cases}$$

One can see from these equations that the rows of the resulting coarse-grained transition generator matrix sum up to zero, and that its off-diagonal elements are positive, which is sufficient to ensure that the coarse-grained transition generator matrix corresponds to a valid Markov process. Once this generator matrix is constructed, the corresponding transition probability matrices for various horizons can be calculated by the standard matrix exponentiation rules [2].

The results of this coarse-graining procedure for the 20-year half-life transition generator and probability matrices, estimated as of 1/1/2005 are shown in Figure 14.

**Figure 14a.**   **Letter-grade transition generator matrix as of 1/1/2005 with 20-year half-life**

| Rating | AAA | AA | A | BAA | BA | B | CCC | D |
|---|---|---|---|---|---|---|---|---|
| AAA | (0.1159) | 0.1134 | 0.0024 | 0.0000 | 0.0000 | 0.0000 | 0.0000 | 0.0000 |
| AA | 0.0311 | (0.1364) | 0.1032 | 0.0020 | 0.0001 | 0.0000 | 0.0000 | 0.0000 |
| A | 0.0001 | 0.0529 | (0.1593) | 0.1024 | 0.0035 | 0.0003 | 0.0000 | 0.0000 |
| BAA | 0.0000 | 0.0007 | 0.0968 | (0.2188) | 0.1113 | 0.0087 | 0.0008 | 0.0004 |
| BA | 0.0000 | 0.0000 | 0.0031 | 0.1206 | (0.3038) | 0.1578 | 0.0153 | 0.0071 |
| B | 0.0000 | 0.0000 | 0.0001 | 0.0040 | 0.0882 | (0.3289) | 0.1617 | 0.0748 |
| CCC | 0.0000 | 0.0000 | 0.0000 | 0.0001 | 0.0030 | 0.0890 | (0.3792) | 0.2870 |
| D | - | - | - | - | - | - | - | - |

**Figure 14b.**   **Letter-grade 1-year transition probability matrix as of 1/1/2005 with 20-year half-life**

| Rating | AAA | AA | A | BAA | BA | B | CCC | D |
|---|---|---|---|---|---|---|---|---|
| AAA | 0.8921 | 0.1002 | 0.0072 | 0.0004 | 0.0000 | 0.0000 | 0.0000 | 0.0000 |
| AA | 0.0274 | 0.8764 | 0.0894 | 0.0062 | 0.0005 | 0.0001 | 0.0000 | 0.0000 |
| A | 0.0008 | 0.0458 | 0.8593 | 0.0854 | 0.0074 | 0.0010 | 0.0002 | 0.0001 |
| BAA | 0.0000 | 0.0028 | 0.0807 | 0.8129 | 0.0867 | 0.0134 | 0.0021 | 0.0014 |
| BA | 0.0000 | 0.0002 | 0.0072 | 0.0938 | 0.7483 | 0.1167 | 0.0201 | 0.0137 |
| B | 0.0000 | 0.0000 | 0.0005 | 0.0071 | 0.0651 | 0.7300 | 0.1146 | 0.0827 |
| CCC | 0.0000 | 0.0000 | 0.0000 | 0.0005 | 0.0050 | 0.0629 | 0.6895 | 0.2421 |
| D | - | - | - | - | - | - | - | 1.0000 |